\documentclass[aps,pre,reprint,showpacs,showkeys,groupedaddress]{revtex4-1}

\usepackage{amsmath}
\usepackage{graphicx}
\usepackage{rotating}
\usepackage{color}
\usepackage{caption}
\usepackage{subcaption}
\newcommand{\rom}[1]{\uppercase\expandafter{\romannumeral #1\relax}}

\begin{document}


\title{Effect of finite range interactions on roton mode softening in a multi-component BEC}

\author{Abhijit Pendse}\thanks{abhijeet.pendse@students.iiserpune.ac.in}
\affiliation{Indian Institute of Science Education and Research, Pune, India 411008.}

\date{\today}

\begin{abstract}
We consider the Gross-Pitaevskii(GP) model of a Bose-Einstein Condensate(BEC) for single-component and multi-component BEC. The pseudopotential for s-wave scattering between atoms is taken to be of width of the order of the s-wave scattering length. Such an interaction giving rise to a roton minimum in the spectrum of elementary excitations of a single component BEC is well known. However, softening roton modes takes us in the strongly interacting BEC regime where three body losses occur. We study the roton mode softening for a multi-component BEC. We show that by increasing the number of components of a multi-component BEC, the roton mode can be softened at a progressively lower value of the gas parameter ($a^{3}n$), thus reducing three body losses.
\end{abstract}


\maketitle


\section{Introduction}

Since the discovery of Bose-Einstein Condensation(BEC) in dilute atomic gases \cite{atomicbec_anderson, atomicbec_ketterle}, there have been many experimental and theoretical investigations in this field.
BEC has been shown to possess superfluidity \cite{ps}, whose phenomenological theory was given by Landau \cite{superfluidity_landau}. An interesting extension of the idea of superfluidity is the supersolid phase, which is superfluidity with solid order \cite{supersolid_leggett, quantumcrystals_chester}. The search for such phases of matter has been a greatly pursued topic and there have been a lot of theoretical proposals in this regards in BEC \cite{roton_rydberg_henkel, roton_softcore_ancilotto}, along with experimental observations \cite{roton_experiment_mottl, roton_blakie}. The conventional approach to obtain such a state in BEC is through softening of roton modes. These have been mainly proposed in dipolar interactions, rydberg dressed atoms and optical lattice mediated interactions in a BEC \cite{roton_santos, roton_rydberg_henkel, roton_softcore_ancilotto}. These approaches take into consideration long range interactions on top of the s-wave scattering between atoms. There have also been studies taking into account the nonlocal nature of s-wave interactions, without taking into account additional long range interactions on top of it \cite{nonlocal_pomeau, nonlocal_pseudopotential_macri}. Occurence and lowering of roton minimum was shown in these cases too.

\par
Apart from changing various parameters in an atomic BEC, like the density and scattering length \cite{feshbach_inouye}, one can also change the number of distinct atomic components. These are the multi-component BECs \cite{twocomponent_cornish, yb_yoshiro, multicomponent_ueda}. The most studied amongst the multi-component BECs are the two-component BECs. There have been extensive studies regarding the miscibility \cite{miscibility_twocomponent_wieman}, stability \cite{stability_twocomponent_savage} and varying the inter-species interaction strength \cite{interspecies_interaction_inguscio} in a two component BEC. BEC with more than two components have also been studied \cite{multicomponent_ueda}. 

\par
 The main tool for theoretical investigations in a BEC has been the mean field Gross-Pitaevskii(GP) equation \cite{gp_gross, pethick}. The local GP equation for a single-component BEC predicts the ground state of the BEC to be one with uniform density \cite{ps}, with the order parameter $\psi(\textbf{r},t)=\sqrt{n} \; e^{-i\mu t/\hbar}$. The quantity $|\psi(\textbf{r},t)|^{2}=n$ gives the density of the condensate. Here $\mu$ is the chemical potential of the BEC. Small amplitude oscillations on top of this uniform density state have the dispersion relation $\omega=\sqrt{(\hbar^{2}k^{4}/(4m^{2}))+(k^{2}gn/m)}$ \cite{pethick}. A similar ground state exists for a multi-component BEC, where all the different components have uniform densities \cite{pethick}. Also, similar dispersion relation exists for excitations on top of the different uniform densities in a multi-component BEC. The point to note here is that, this dispersion is in the presence of the $\delta$-function pseudopotential, i.e. one whose range tends to zero. 

\par
For an atomic BEC with temperature of the order of few hundred nano-Kelvin, the atoms have very low momenta. Hence, the details of the inter-atomic potential can be ignored and the real potential can be replaced by an effective pseudopotential(say $V_{eff}$) which is repulsive, for the symmetric s-wave scattering between atoms \cite{ps}. The condition for using this approximation is that $\int\textbf{dr}\;{V_{eff}}=g$ holds, where $g=4\pi\hbar^{2}a/m$ denotes the strength of the inter-atomic s-wave interactions. This pseudopotential is taken to be a $\delta$-function in a dilute BEC. However, instead of taking a $\delta$-function, one may take a repulsive rectangular barrier as the scattering pseudopotential which may be able to capture the non-local behaviour of the inter-particle interactions. This sort of pseudopotential has been considered in literature \cite{nonlocal_ketterle, nonlocal_kutz, nonlocal_pomeau}. This sort of extended pseudopotential gives rise to the formation of a roton minimum in a single component BEC \cite{nonlocal_pomeau, nonlocal_pseudopotential_macri, roton_softcore_ancilotto}. This roton minimum can be further lowered by varying the interaction strength between atoms.

\par
In this article, we study the effect of taking a non-local pseudopotential for s-wave scattering on the dispersion relation of a multi-component BEC. We consider the nonlocal s-wave scattering and do not consider any additional long range interactions. We start by discussing the existing results for small amplitude excitations on top of uniform density for single component BECs with $\delta$-function pseudopotential. We then briefly discuss the extended pseudopotential in the form of a rectangular barrier which is extensively studied in literature \cite{nonlocal_pomeau, nonlocal_pseudopotential_macri, roton_softcore_ancilotto, nonlocal_pethick}. This barrier pseudopotential is a simple nonlocal extension of the $\delta$-function pseudopotential. We use this barrier pseudopotential to study elementary excitations in single as well as multi-component BEC. 

\par

The use of the delta function pseudopotential gives good agreement with the experiments for dilute gases where $a^{3}n<<1$. However, when we go into the regimes where $a^{3}n\rightarrow 1$, one has to consider corrections on top of the GP equation with delta function pseudopotential \cite{bragg_strongly_interacting_papp, tunability_strongly_interacting_pollack} to fully characterize the properties of the system. This is mainly because as $a^{3}n\rightarrow 1$, the finite range effects of the BEC particle interactions come into the picture. A simple way to account for these finite range effects is to consider an extended pseudopotential in place of $\delta$-function pseudopotential. This is exactly what we do in this paper. We would like to re-emphasize on the fact that the extended pseudopotential used by us is, as the name suggests, {\textit{a pseudopotential}} and not the actual interaction potential between the BEC particles. As mentioned before, such an approach has already been considered for single component BEC in literature \cite{nonlocal_pomeau, nonlocal_pseudopotential_macri, roton_softcore_ancilotto}.

\par
We first revisit the roton minimum appearing in the elementary excitation spectrum of a single component BEC with extended pseudopotential which has been studied previously \cite{nonlocal_pomeau, nonlocal_pseudopotential_macri}. This roton minimum can be lowered by increasing the gas parameter $a^{3}n$. However, this appearance and lowering of the roton minimum lies well outside the diluteness limit, and in the strongly interacting regime, for a single-component BEC. The diluteness limit is given by $a^{3}n<<1$ for a single component BEC. A similar diluteness limit exists for a multi-component BEC given by $\sqrt{n_{i}n_{j}}a_{ij}^{3}<<1$ \cite{diluteness_yukalov}, where $n_{i}$ and $n_{j}$ are the densities of the $i^{th}$ and $j^{th}$ components and $a_{ij}$ is the inter-species scattering length. As we go on going away from the diluteness limit and towards the strongly interacting regime $a^{3}n\sim 1$, the lifetime of a BEC decreases due to the appearance of three body effects. Hence, it is important to achieve roton mode softening at a lower value of gas parameter, so that we may observe it over a significant interval of time. With this motivation, we study the effect of non-locality of s-wave scattering on the appearance and lowering of the roton minimum for a BEC with multiple components. We investigate whether we can soften the roton minimum in a multi-component BEC at a lower value of the gas parameter $\sqrt{n_{i}n_{j}}a_{ij}^{3}$ as compared to that for a single-component BEC. We show that for multi-component BECs, the appearance and lowering of roton minimum can be achieved at a lower value of the gas parameter as we go on increasing the number of distinct components. We observe that even for a two-component BEC, the roton mode softening can be achieved at a significantly lower value of the gas parameter.



\section{Single component BEC}

\subsection{Excitations with local interactions}
Let us discuss briefly the dispersion relation for a single component BEC with $\delta$-function inter-particle pseudopotential given by $V_{eff}(\textbf{r}-\textbf{r}')=g\;\delta(\textbf{r}-\textbf{r}')$. This is a standard textbook calculation \cite{ps}. In the absence of an external potential, the Gross-Pitaveskii(GP) equation with such contact interactions is given by

\begin{equation}
\label{eq:contact_gp}
i\hbar\frac{\partial \psi({\bf{r}},t)}{\partial t}=-\frac{\hbar^{2}}{2m}\nabla^{2}\psi({\bf{r}},t)+g|\psi(\bf{r},t)|^{2}\psi({\bf{r}},t).
\end{equation}

This equation admits a solution with uniform density. This uniform density state of a BEC is given by $\psi({\bf{r}},t)=\sqrt{n}e^{-i\mu t/\hbar}$, where $|\psi({\bf r},t)|^{2}$ gives the uniform density and $\mu$ is the chemical potential. Small amplitude oscillations on top of this uniform density ground state are taken to be of the form $\theta({\bf r},t)=\sum_l{[u_l({\bf r})e^{\frac{-i\omega_l t}{\hbar}}+v^*_l({\bf r})e^{\frac{i\omega_l t}{\hbar}}]}e^{-i\mu t/\hbar}$. The calculation of the small amplitude oscillations over a uniform density state is a standard procedure\cite{ps} which involves treating $u_{l}$ and $v_{l}$ as excitations with small amplitudes and hence neglecting their higher powers. Now, we drop the subscript $l$ and take $u({\bf{r}})=u e^{i{\bf{k.r}}}$, $v({\bf{r}})=v e^{i{\bf{k.r}}}$ and collect terms which evolve in time identically. This procedure gives us the dispersion relation for excitations as $\omega=\sqrt{(\hbar^{2}k^{4}/(4m^{2}))+(k^{2}gn/m)}$. This relation tells us that for small $k$, the dispersion is phonon like, whereas for large $k$ it is particle like. To keep in mind is the fact that the current approach uses the $\delta$-function approximation for the pseudopotential, which is equivalent to taking the range of a rectangular barrier pseudopotential tending to zero.

\par
 Next we briefly describe the extended pseudopotential used in literature and then use it to study the dispersion relation for a single component and then multi-component BECs in the presence of non-local s-wave scattering.


\subsection{Extended pseudopotential}

In an atomic BEC, due to the low temperatures involved, we are mainly concerned with the low energy scattering between atoms. The most important contribution would then come from the s-wave scattering. So, instead of taking the actual interaction potential between atoms, we can work with a pseudopotential($V_{eff}$), with the condition that $\int\textbf{dr}\;{V_{eff}}=g$, where $g=4\pi\hbar^{2}a/m$, $a$ being the s-wave scattering length\cite{ps}. The standard pseudopotential used is the $\delta$-function pseudopotential given by $V_{eff}=g\;\delta({\bf r}-{\bf r}')$. However, there have been many works which use an extended pseudopotential in the form of a rectangular barrier \cite{roton_softcore_ancilotto, nonlocal_pomeau, nonlocal_ketterle}. This is the simplest extension of the $\delta$-function pseudopotential to incorporate the effects of non-locality of inter-atomic interactions. Pomeau \textit{et al.} \cite{nonlocal_pomeau} have taken the width of this rectangular barrier pseudopotential to be the s-wave scattering length. This is justified since, while considering low energy scattering from a rectangular barrier, the width of the barrier is of the order of the s-wave scattering length( see for example, book by Sakurai \cite{sakurai}). Considering such an extended pseudopotential leads to formation of a roton minimum in the spectrum of elementary excitations which can be further lowered by increasing the parameter $a^{3}n$. In what follows, we use this barrier pseudopotential and study its effect on the excitation spectrum of single as well as multi-component BEC. We take the width of the barrier pseudopotential to be of the order of the s-wave scattering length. We would like to emphasize here that we do not state that the width of the pseudopotential is exactly equal to the s-wave scattering length, but it is rather of the order of the s-wave scattering length.


\subsection{Excitations with non-local interactions}

Let us now look at the effect of non-local interactions on the small amplitude oscillation modes on top of the uniform density state of a BEC. As mentioned before, the uniform density state of a BEC is given by $\psi({\bf{r}},t)=\sqrt{n}e^{-i\mu t/\hbar}$. Let us take the non-local interaction pseudopotential as $V_{eff}({\bf r-r}')=3g/[4\pi(\sigma a)^{3}]$ for $|{\bf r-r}'| \leq \sigma a$ and zero otherwise, where $\sigma$ is an order unity proportionality factor. Thus the width of the scattering pseudopotential is of the order of the s-wave scattering length.

\par
The non-local GP equation in presence of the pseudopotential mentioned above can be derived from the full non-local GP equation, given by \cite{ps}

\begin{equation*}
\begin{split}
i\hbar\frac{\partial \psi({\bf{r}},t)}{\partial t}&=-\frac{\hbar^{2}}{2m}\nabla^{2}\psi({\bf{r}},t)\\
&+\psi({\bf{r}},t)\int{\psi^{*}({\bf r}',t) V_{eff}({\bf r-r}') \psi({\bf r}',t) d{\bf r}'},
\end{split}
\end{equation*}

Inserting the pseudopotential mentioned above, we get,

\begin{equation}
\label{eq:nonlocal_gp_3d}
\begin{split}
i\hbar\frac{\partial \psi({\bf{r}},t)}{\partial t}&=-\frac{\hbar^{2}}{2m}\nabla^{2}\psi({\bf{r}},t)\\
&+\frac{3g}{(\sigma a)^{3}}\psi({\bf{r}},t)\int_{0}^{\sigma a}{|\psi({\boldsymbol{\phi}},t)|^2 d{\boldsymbol{\phi}}},
\end{split}
\end{equation}

where, ${\boldsymbol {\phi}}={\bf r-r}'$ and $d\boldsymbol{\phi}$ represents the infinitesimal volume element. The integral limits are so taken since we have taken the width of the scattering barrier to be of the order of the s-wave scattering length. We further set the dimensionless constant $\sigma$ to unity for brevity. This we do while mentioning the caveat that the range of the pseudopotential is not exactly the scattering length, but is of the order of the scattering length.

\par
Once again, we take the form of excitations on top of the uniform density state as $\theta({\bf r},t)=\sum_l{[u_l({\bf r})e^{\frac{-i\omega_l t}{\hbar}}+v^*_l({\bf r})e^{\frac{i\omega_l t}{\hbar}}]}e^{-i\mu t/\hbar}$ and treat $u$ and $v$ as excitations with small amplitudes and hence neglect their higher powers. As for the local interactions case, taking $u(\textbf{r})=u e^{i\textbf{k.r}}$, $v(\textbf{r})=v e^{i\textbf{k.r}}$ and collecting terms which evolve in time identically, we get

\begin{equation*}
\omega=\pm \frac{\hbar}{m} \sqrt{\frac{k^{4}}{4}+(4\pi k n) \:\sin{\sigma ka}}.
\end{equation*}

which can be equivalently written as

\begin{equation}
\label{eq:single_component_roton}
\omega=\pm \frac{\hbar}{a^{2}m} \sqrt{\frac{\lambda^{4}}{4}+(4\pi a^{3} n\lambda) \:\sin{\sigma\lambda}},
\end{equation}

where $\lambda=ka$. The gas parameter $a^{3}n$ appearing in the equation above should be small in order for the three body interactions to be noticeably small. Particularly, the well known diluteness limit is given by $a^{3}n<<1$. However, there have been recent experiments which have explored the limits of the gas parameter, trying to push it towards the strongly interacting regime $a^{3}n \sim 1$ \cite{bragg_strongly_interacting_papp, tunability_strongly_interacting_pollack}.

\begin{figure}[h]
\rotatebox{270}  { \includegraphics[width=6cm]{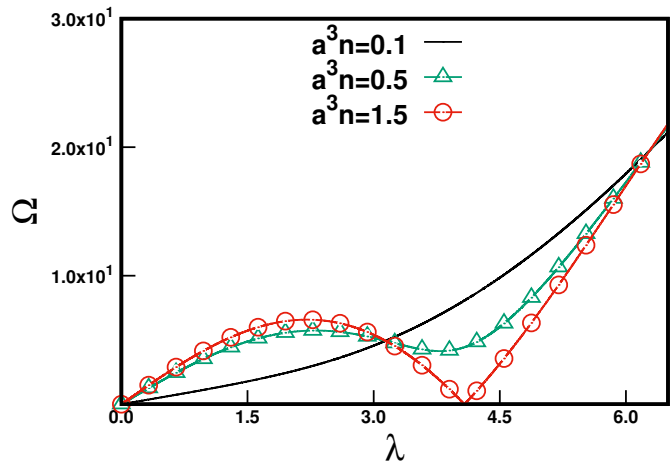}  }
   \caption{Figure shows the dispersion relation for single component BEC in the presence of non-local inter-particle interaction pseudopotential for certain values of the parameter $a^{3}n$. Here, $\Omega=m\omega a^{2}/\hbar$ and $\sigma=1$ for the sake of comparison.}
\label{fig:single}
\end{figure}

The dispersion relation given by Eq.(\ref{eq:single_component_roton}) exhibits a roton minimum which can be lowered as we increase $a^{3}n$ as shown in Fig.(\ref{fig:single}). The roton minimum touches the $ak$ axis at a value of $a^{3}n\sim1.5$ which is beyond the $a^{3}n<<1$ limit. This would take us well into the strongly interacting regime. Nonetheless, we see that it is possible to pull the roton minimum down in principle. Thus we get a hint that it might be possible to pull the roton minimum down, at a lower value of $a^{3}n$, by introducing additional interactions such as those in multi-component BEC. Since the macroscopic wave-function of different components in a multi-component BEC is different, there occur cross terms between the wave-functions in the GP equation, due to the inter-species scattering. Thus for multi-component BEC, there exist additional interaction terms than the ones in the GP equation for a single component BEC. In the following section, we consider multi-component BEC and study the manifestation of inter-species interactions in the dispersion relation of small amplitude oscillations.

\section{Multi-component BEC}

\subsection{Excitations with local interactions}

BECs with more than one components, in the absence of an external potential, can be analysed by taking into account the inter-particle interaction in the GP energy functional. This can be written as \cite{thesis_pattinson}

\begin{widetext}
\begin{equation}
\label{eq:multi_energy}
\begin{split}
E&=\frac{\hbar^{2}}{2m}\sum_{j}\int{{\bf dr}\;|\nabla \psi_{j}({\bf r})|^{2}}+\sum_{j}\int{{\bf dr}\;\frac{|\psi_j({\bf r})|^{2}}{2}\int{{\bf dr^{'}}\psi_{j}^{*}({\bf r^{'}}) V_{jj}({\bf r-r^{'}})\psi_{j}({\bf r^{'}})}} \\
&+\sum^{j>l}_{j,l}\int{\int{{\bf dr^{'}}{\bf dr}\;\Big(\frac{\psi_{j}({\bf r})\psi_{j}^{*}({\bf r}) V_{jl}({\bf r-r^{'}})\psi_{l}({\bf r^{'}})\psi_{l}^{*}({\bf r^{'}})}{2}\Big)}}.
\end{split}
\end{equation}
\end{widetext}

where $g_{ij}$ represents the inter-species s-wave scattering strength, $g_{ij}=4\pi\hbar^{2}a_{ij}(m_{i}+m_{j})/(m_{i}m_{j})$, $m_{i}$ and $m_{j}$ represent the mass of the atoms of different species whose wavefunctions are $\psi_{i}$ and $\psi_{j}$ respectively. The inter-particle scattering length is given by $a_{ij}$. If the scattering pseudopotential is taken to be $\delta$-function as before with $V^{eff}_{ii}({\bf r-r}')=g_{ii}\delta({\bf r-r}')$ for intra-species scattering and $V^{eff}_{ij}({\bf r-r}')=g_{ij}\delta({\bf r-r}')$, we get the local GP equation for multi-component BEC as

\begin{widetext}
\begin{equation}
\label{eq:multi_gp_local}
i\hbar\frac{\partial \psi_{j}({\bf r},t)}{\partial t}=-\frac{\hbar^{2}}{2m}\nabla^{2}\psi_{j}({\bf r},t)+g_{jj}|\psi_{j}({\bf r},t)|^{2}\psi_{j}({\bf r},t)+\psi_{j}({\bf r},t)\sum_{l}^{j\neq l}g_{jl}|\psi_{l}({\bf r},t)|^{2},
\end{equation}
\end{widetext}

for each of the several components, denoted by the subscript $j$.

Now, as for the single component case, these equations admit solutions with uniform density for each component given by $\psi_{j}(\textbf{r},t)=\sqrt{n_{j}}e^{-i\mu_{j}t/\hbar}$. Such BEC might exist stably in many configurations with regards to the miscibility. The general miscibility condition for two components $i$ and $j$ is given by $a_{ij}<\sqrt{a_{i}a_{j}}$ \cite{miscibility_ao}. We shall consider here a uniform, homogeneous mixture state of the many components, i.e. we consider a state where the miscibility criterion holds for all components.
We shall first look at the dispersion relation of small amplitude oscillations for the local GP equation given by Eq.(\ref{eq:multi_gp_local}) and then look at the non-local case in the next section.

\par
We take the small amplitude excitations over the state $\psi_{j}(\textbf{r},t)=\sqrt{n_{j}}e^{-i\mu_{j}t/\hbar}$ of the form  $\theta_{j}({\bf r},t)=\sum_q{[u_{qj}({\bf r})e^{\frac{-iw_{q} t}{\hbar}}+v^*_{qj}({\bf r})e^{\frac{iw_{q} t}{\hbar}}]}e^{-i\mu_{j} t/\hbar}$. Further, as for the single component case, we take the form of $u_{j}$ and $v_{j}$ as  $u_{qj}(\textbf{r})=u_{qj} e^{i\bf{k_{q}.r}}$, $v_{qj}(\textbf{r})=v_{qj} e^{i\bf{k_{q}.r}}$. For a two component BEC, this dispersion relation has two distinct branches \cite{two_component_excitations_alexandrov}. In general for a multicomponent BEC we may get a complicated dispersion relation for excitations. To simplify the system, we assume that all the intra-species scattering lengths are equal($a_{1}$ say), so are the inter-species scattering lengths($a_{2}$ say). This simplification enables us to look particularly into the effects of the presence of multiple components in a BEC on the roton mode softening. This simplification also brings out the role of inter-species interaction on roton mode softening, which is our core interest here.
\par
 Taking species with equal densities($n$) and comparable masses($m$) we get the following matrix structure for excitations

\begin{widetext}
\begin{equation}
\label{eq:matrix_local}
\begin{pmatrix} 
{U_{-}} & {X} & {c} & {c}&{........}&{}&{}&{} \\ 
{X} & {U_{+}}  & {c} & {c}&{........}&{}&{}&{}\\
{c} & {c}&{U_{-}} & {X} & {c} & {c}&{....}&{} \\
{c}&{c}&{X} & {U_{+}}  & {c} & {c}&{....}&{}\\
{}&{}&{}&{.}&{}&{}&{}&{}\\
{}&{}&{}&{}&{.}&{}&{}&{}\\
{}&{}&{}&{}&{}&{.}&{}&{}\\
{}&{}&{}&{}&{}&{}&{.}&{}\\
{}&{}&{}&{........}&{c}&{c}&{U_{-}}&{X}\\
{}&{}&{}&{........}&{c}&{c}&{X}&{U_{+}}\\
\end{pmatrix}
\begin{pmatrix}
u_1 \\ v_1 \\.\\.\\.\\. \\.\\.\\ u_{s} \\ v_{s}
\end{pmatrix}
=
\begin{pmatrix}
0 \\ 0 \\.\\.\\.\\.\\.\\.\\0\\0
\end{pmatrix}
,
\end{equation}
where

\begin{equation*}
\begin{split}
U_{\pm}:=&\pm w\hbar-\frac{\hbar^{2}k^{2}}{2m}-g_{1}n \\
X:=&-g_{1}n\\
c:=&-g_{2}n,
\end{split}
\end{equation*}
\end{widetext}

\par
with $s$ as the total number of distinct components in the multi-component BEC.

 To get a non-trivial solution for excitations $u$ and $v$ on top of the uniform density ground state, the determinant of the $2s\times 2s$ square matrix in Eq.(\ref{eq:matrix_local}) must be zero. This condition gives us

\begin{widetext}
\begin{equation*}
\Big[w^{2}-\frac{\hbar^{2}}{m^{2}}\Big(\frac{k^{4}}{4}+4\pi na_{1}k^{2}+(s-1)4\pi na_{2}k^{2}\Big)\Big] \Big[w^{2}-\frac{\hbar^{2}}{m^{2}}\Big(\frac{k^{4}}{4}+4\pi na_{1}k^{2}-4\pi na_{2}k^{2}\Big)\Big]^{(s-1)}=0.
\end{equation*}
\end{widetext}

This equation gives us two branches of the dispersion relation as,

\begin{equation}
\label{eq:excitations_local}
\begin{split}
w=&\frac{\hbar}{m}\Big[\frac{k^{4}}{4}+4\pi na_{1}k^{2}+(s-1)4\pi na_{2}k^{2}\Big]^{\frac{1}{2}}\\
& \\
& or/and\\
& \\
w=&\frac{\hbar}{m}\Big[\frac{k^{4}}{4}+4\pi na_{1}k^{2}-4\pi na_{2}k^{2}\Big]^{\frac{1}{2}}
\end{split}
\end{equation}

\vspace{5mm}
\subsection{Excitations with non-local interactions}

Let us now consider non-local pseudopotential for both intra-species and inter-species interactions in a multi-component BEC. Let the interaction pseudopotential be given by $V^{eff}_{jj}({\bf r-r}')=3g_{jj}/4\pi(\sigma a_{jj})^{3}$ for $|{\bf r-r}'| \leq \sigma a_{jj}$ and zero otherwise for the intra-species interaction. For the inter-species interaction, we take the interaction psuedopotential as $V^{eff}_{ij}({\bf r-r}')=3g_{ij}/4\pi(\sigma a_{ij})^{3}$ for $ |{\bf r-r}'| \leq \sigma a_{ij}$ and zero otherwise. We would like to re-emphasize the fact that we do not claim the range of the pseudopotential to be exactly equal to that of the scattering length, but it is of the order of the scattering length, which is what is considered here to simplify the model.

\par
Now, we write the non-local GP equation which can be derived from the energy functional in Eq.(\ref{eq:multi_energy}) by considering the interaction pseudopotential to be non-local, as given above. It is of the form

\begin{widetext}
\begin{equation}
\label{eq:multi_gp_nonlocal}
i\hbar\frac{\partial \psi_{j}({\bf r},t)}{\partial t}=-\frac{\hbar^{2}}{2m}\nabla^{2}\psi_{j}({\bf r},t)+\frac{3g_{jj}}{(\sigma a_{jj})^{3}}\psi_{j}({\bf r},t)\int_{0}^{\sigma a_{jj}}{|\psi_{j}(\boldsymbol{\phi},t)|^2 d\boldsymbol{\phi}}+\sum^{j\neq l}_{l}\Big[\frac{3g_{jl}}{(\sigma a_{jl})^{3}}\psi_{j}({\bf r},t)\int_{0}^{\sigma a_{jl}}{\psi_{l}^{*}(\boldsymbol{\phi},t)\cdot \psi_{l}(\boldsymbol{\phi},t) d\boldsymbol{\phi}} \Big].
\end{equation}
\end{widetext}

As for the single component case,  $\boldsymbol{\phi}={\bf r-r}'$ and $d\boldsymbol{\phi}$ represents the infinitesimal volume element. The above equation admits a solution where each of the species have a uniform density, same as that for the local case. This can be verified by plugging in the form for different $\psi_{j}$ as $\psi_{j}({\bf r},t)=\sqrt{n_{j}}e^{-i\mu_{j}t/\hbar}$. As before, considering small amplitude oscillations $\theta_{j}({\bf r},t)$ on top of this uniform density state, we get the following matrix by mode matching.

\begin{widetext}
\begin{equation}
\label{eq:matrix}
\begin{pmatrix} 
{A_{-}} & {B} & {D} & {D}&{........}&{}&{}&{} \\ 
{B} & {A_{+}}  & {D} & {D}&{........}&{}&{}&{}\\
{D} & {D}&{A_{-}} & {B} & {D} & {D}&{....}&{} \\
{D}&{D}&{B} & {A_{+}}  & {D} & {D}&{....}&{}\\
{}&{}&{}&{.}&{}&{}&{}&{}\\
{}&{}&{}&{}&{.}&{}&{}&{}\\
{}&{}&{}&{}&{}&{.}&{}&{}\\
{}&{}&{}&{}&{}&{}&{.}&{}\\
{}&{}&{}&{........}&{D}&{D}&{A_{-}}&{B}\\
{}&{}&{}&{........}&{D}&{D}&{B}&{A_{+}}\\
\end{pmatrix}
\begin{pmatrix}
u_1 \\ v_1 \\.\\.\\.\\. \\.\\.\\ u_{s} \\ v_{s}
\end{pmatrix}
=
\begin{pmatrix}
0 \\ 0 \\.\\.\\.\\.\\.\\.\\0\\0
\end{pmatrix}
,
\end{equation}
where

\begin{equation*}
\begin{split}
A_{\pm}:=&\pm\omega\hbar-\frac{\hbar^{2}k^{2}}{2m}-g_{1}n\frac{\sin{\sigma a_{1}k}}{a_{1}k} \\
B:=&-g_{1}n\frac{\sin{\sigma a_{1}k}}{a_{1}k}\\
D:=&-g_{2}n\frac{\sin{\sigma a_{2}k}}{a_{2}k}.
\end{split}
\end{equation*}
\end{widetext}
 Here too we have considered the simplification that all intra-species scattering lengths are equal($a_{1}$) and so are the inter-species scattering lengths($a_{2}$).

As before, to get a non-trivial solution for excitations $u$ and $v$ on top of the uniform density ground state, the determinant of the $2s\times 2s$ square matrix in Eq.(\ref{eq:matrix}) must be zero. Since the matrix structure is similar to the one for the local multi-component case, we get the following equation

\begin{widetext}
\begin{equation*}
\begin{split}
&\Big[\omega^{2}-\frac{\hbar^{2}}{m^{2}}\Big(\frac{k^{4}}{4}+4\pi na_{1}k^{2}\Big(\frac{\sin{\sigma a_{1}k}}{a_{1}k}\Big)+(s-1)4\pi na_{2}k^{2}\Big(\frac{\sin{\sigma a_{2}k}}{a_{2}k}\Big)\Big)\Big]\\
&\times \Big[\omega^{2}-\frac{\hbar^{2}}{m^{2}}\Big(\frac{k^{4}}{4}+4\pi na_{1}k^{2}\Big(\frac{\sin{\sigma a_{1}k}}{a_{1}k}\Big)-4\pi na_{2}k^{2}\Big(\frac{\sin{\sigma a_{2}k}}{a_{2}k}\Big)\Big)\Big]^{(s-1)}=0.
\end{split}
\end{equation*}
\end{widetext}

 This condition gives us two branches of the dispersion relation for excitations as

\begin{widetext}
\begin{equation}
\label{eq:multi_roton}
\begin{split}
\omega=&\frac{\hbar}{m}\Big[\frac{k^{4}}{4}+4\pi na_{1}k^{2}\Big(\frac{\sin{\sigma a_{1}k}}{a_{1}k}\Big)+(s-1)4\pi na_{2}k^{2}\Big(\frac{\sin{\sigma a_{2}k}}{a_{2}k}\Big)\Big]^{\frac{1}{2}}\\
&\\
& or/and\\
&\\
\omega=&\frac{\hbar}{m}\Big[\frac{k^{4}}{4}+4\pi na_{1}k^{2}\Big(\frac{\sin{\sigma a_{1}k}}{a_{1}k}\Big)-4\pi na_{2}k^{2}\Big(\frac{\sin{\sigma a_{2}k}}{a_{2}k}\Big)\Big]^{\frac{1}{2}}.
\end{split}
\end{equation}
\end{widetext}

The above equations exhibit a roton minimum, similar to the one for a single component BEC with non-local pseudopotential. However, there is a crucial difference, since now the factor $s$ in the above equation means that the dispersion relation would depend on the number of distinct species in the multi-component BEC. Also, for a single component BEC there was just one length scale for the scattering length. However, for a multi-component BEC, on top of the intra-species scattering length there exist inter-species scattering length as well.

\begin{figure*}
\captionsetup[subfigure]{oneside,margin={1.0cm,-2.5cm}}
    \begin{subfigure}[b]{0.2\textwidth}
\rotatebox{270}{        \includegraphics[width=\textwidth]{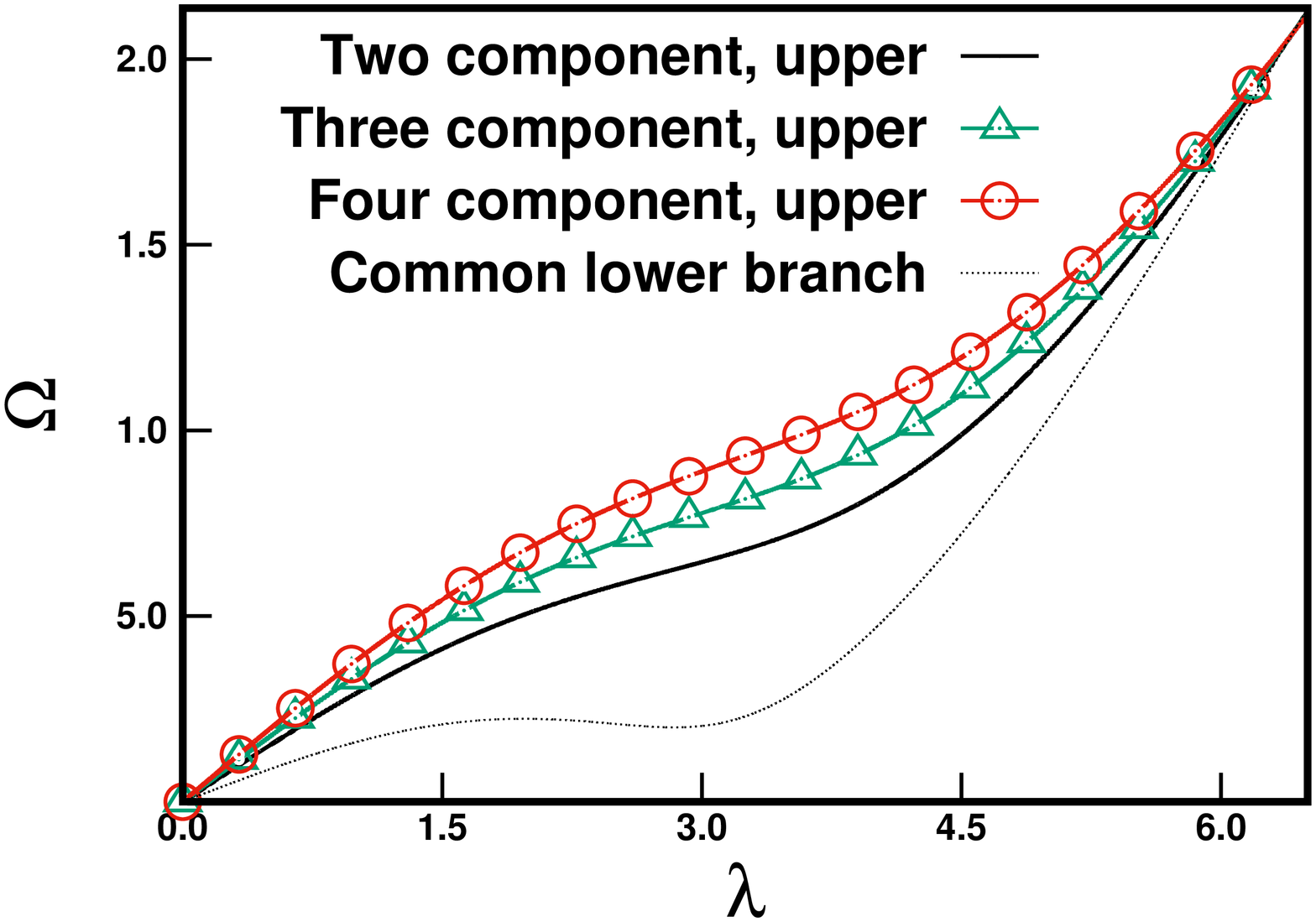}}      
   \caption{$\beta=0.49$, $a_{1}^{3}n=0.5$}
       \label{fig:lower_11}
    \end{subfigure}
\hfill%
    ~ 
   \begin{subfigure}[b]{0.2\textwidth}
\rotatebox{270}{        \includegraphics[width=\textwidth]{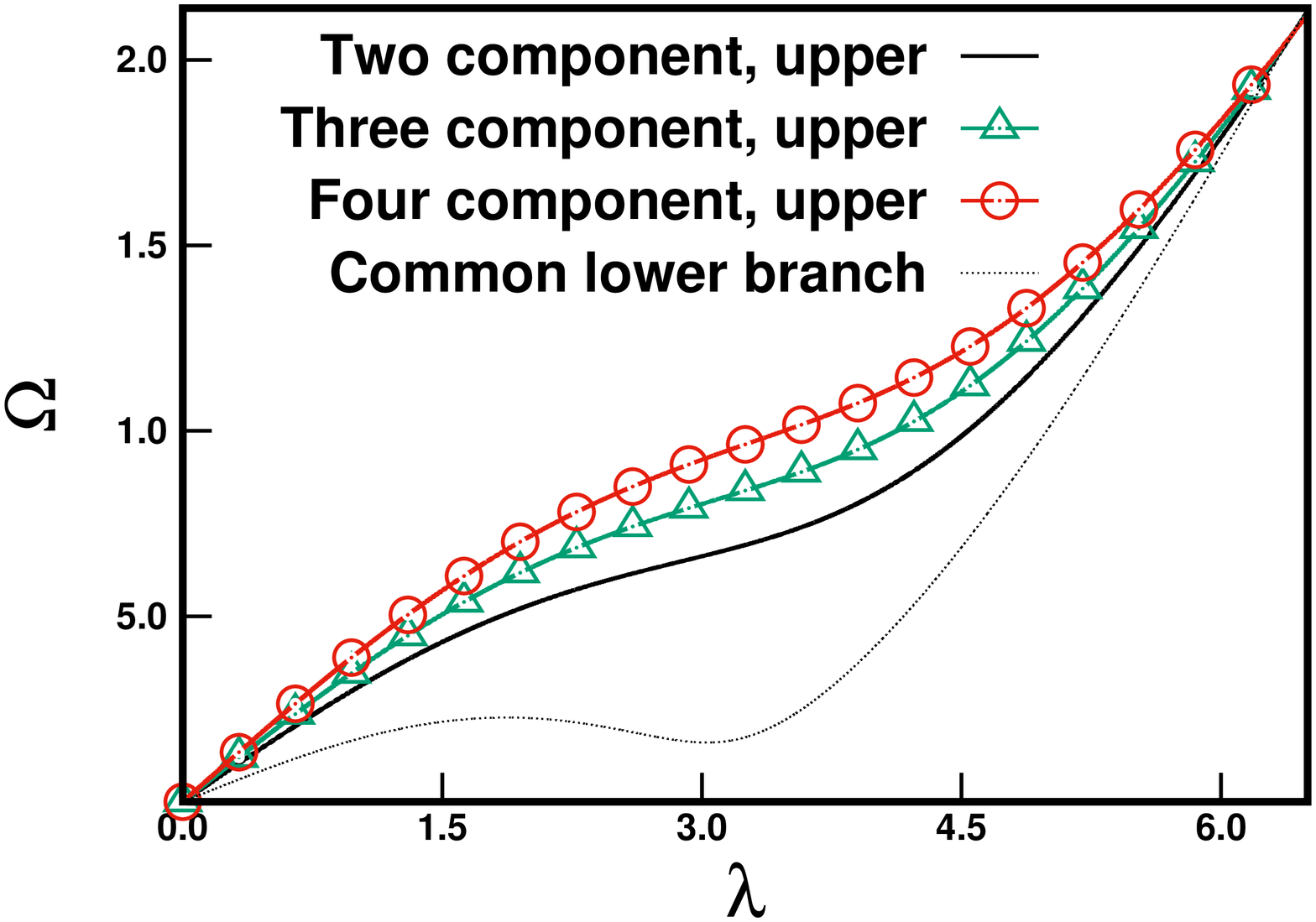}}
        \caption{$\beta=0.49$, $a_{1}^{3}n=0.55$}
        \label{fig:lower_12}
    \end{subfigure}
\hfill
 ~ 
 \begin{subfigure}[b]{0.2\textwidth}
\rotatebox{270}       { \includegraphics[width=\textwidth]{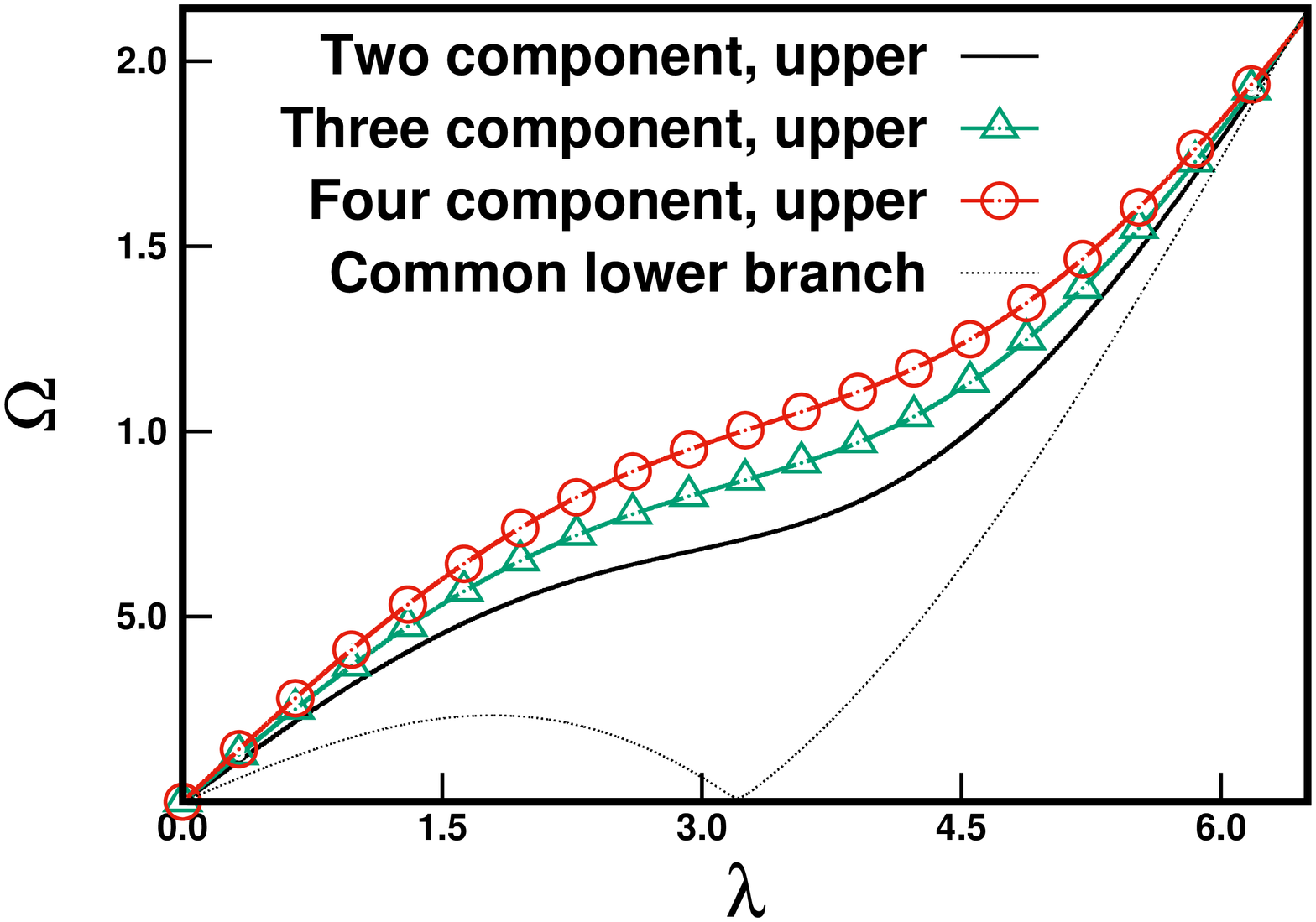}}
        \caption{$\beta=0.49$, $a_{1}^{3}n=0.616$}
        \label{fig:lower_13}
    \end{subfigure}
\hspace*{\fill}

    \caption{Figures shows the dispersion relation for small amplitude excitations of a BEC, in the presence of non-local interaction pseudopotential between atoms, for the common branch of elementary excitations for multi-component BEC. The plots are for different values of the intra-species gas parameter $a_{1}^{3}n$. Here, $\Omega=m\omega a_{1}^{2}/\hbar$ and $\sigma=1$ for the sake of comparison.}
\label{fig:lower}
\end{figure*}

\begin{figure*}
\captionsetup[subfigure]{oneside,margin={1.0cm,-2.5cm}}
    \begin{subfigure}[b]{0.2\textwidth}
\rotatebox{270}{        \includegraphics[width=\textwidth]{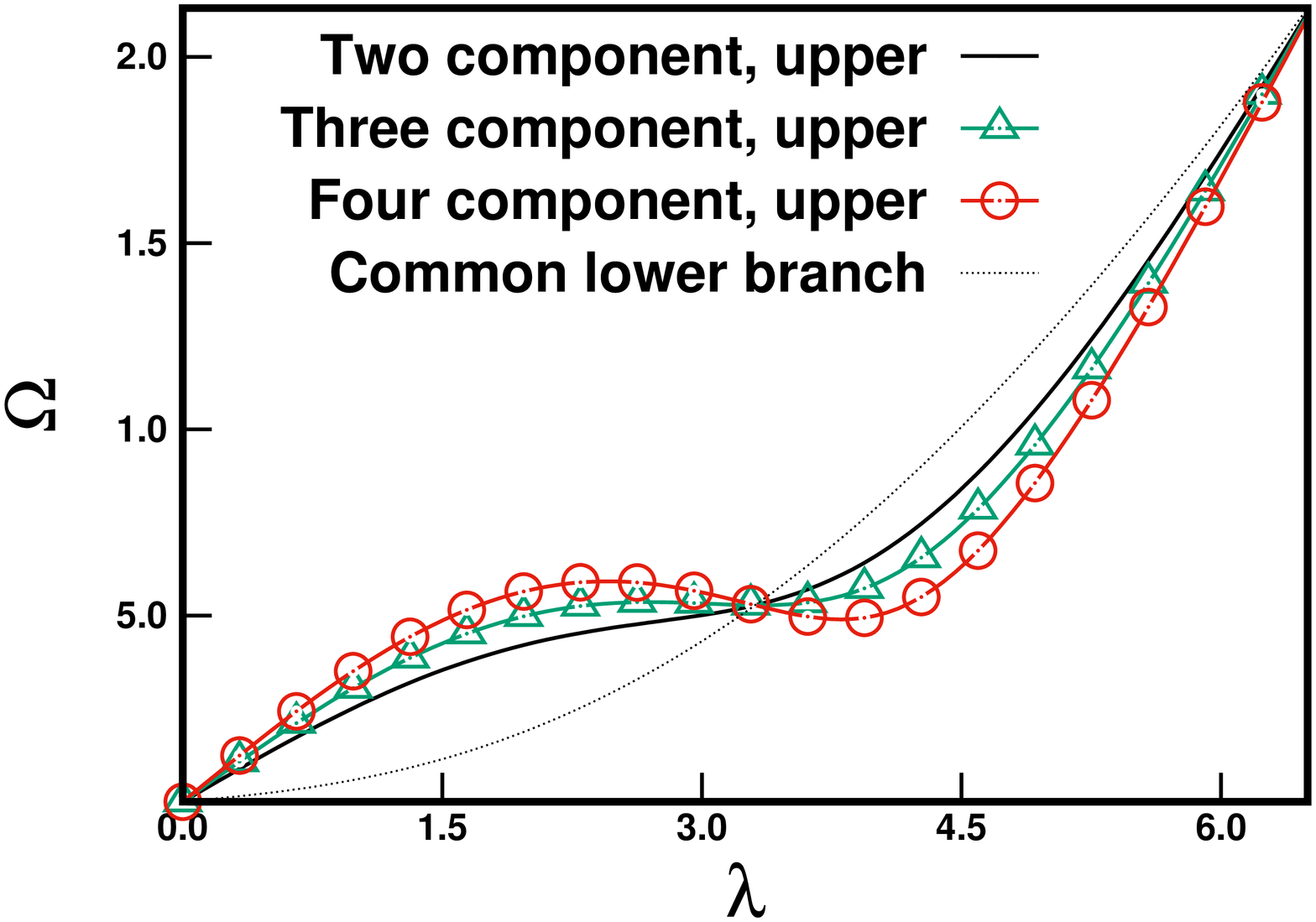}}      
   \caption{$\beta=0.95$, $a_{1}^{3}n=0.3$}
       \label{fig:multi_11}
    \end{subfigure}
\hfill%
    ~ 
   \begin{subfigure}[b]{0.2\textwidth}
\rotatebox{270}{        \includegraphics[width=\textwidth]{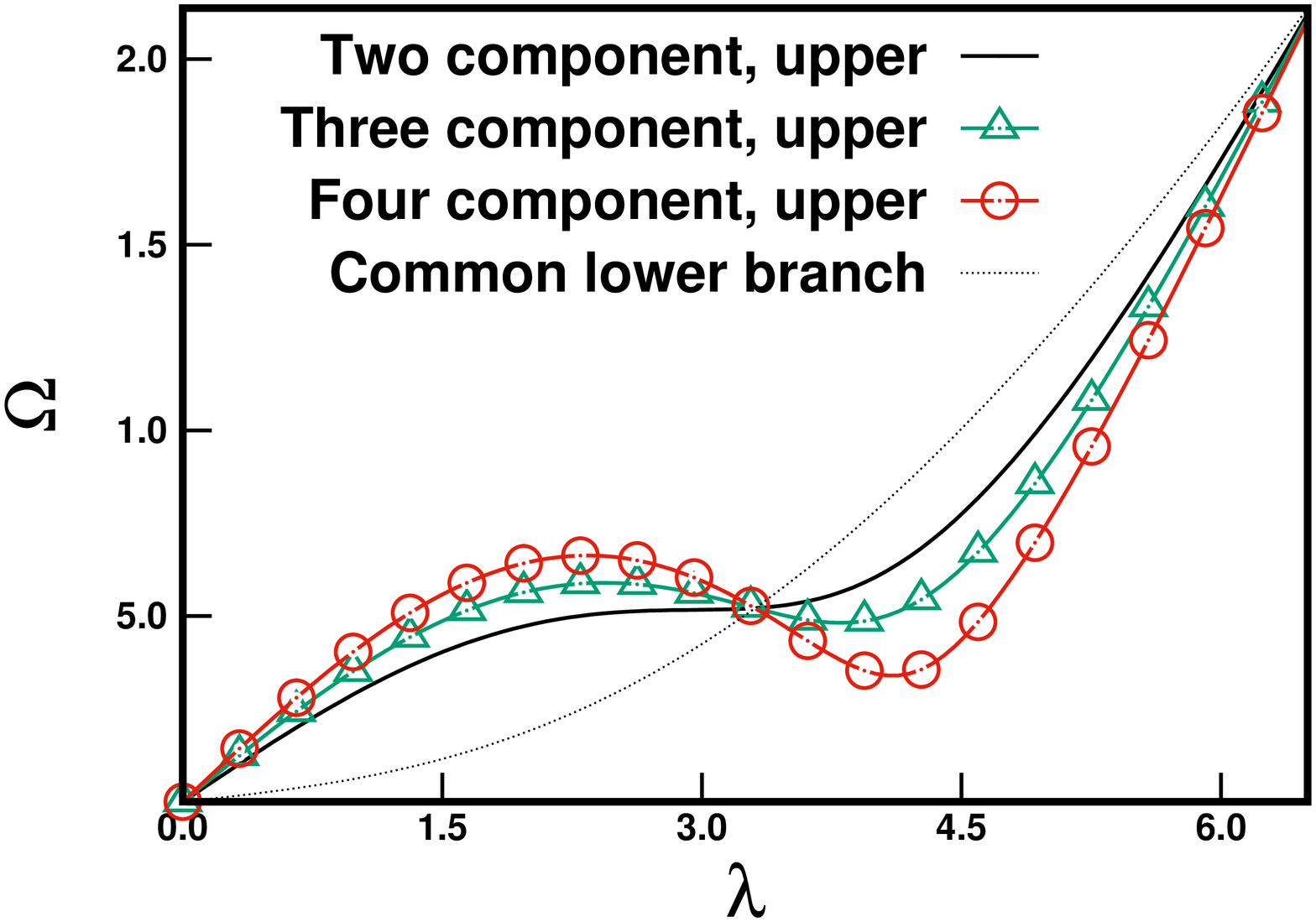}}
        \caption{$\beta=0.95$, $a_{1}^{3}n=0.4$}
        \label{fig:multi_12}
    \end{subfigure}
\hfill
 ~ 
 \begin{subfigure}[b]{0.2\textwidth}
\rotatebox{270}       { \includegraphics[width=\textwidth]{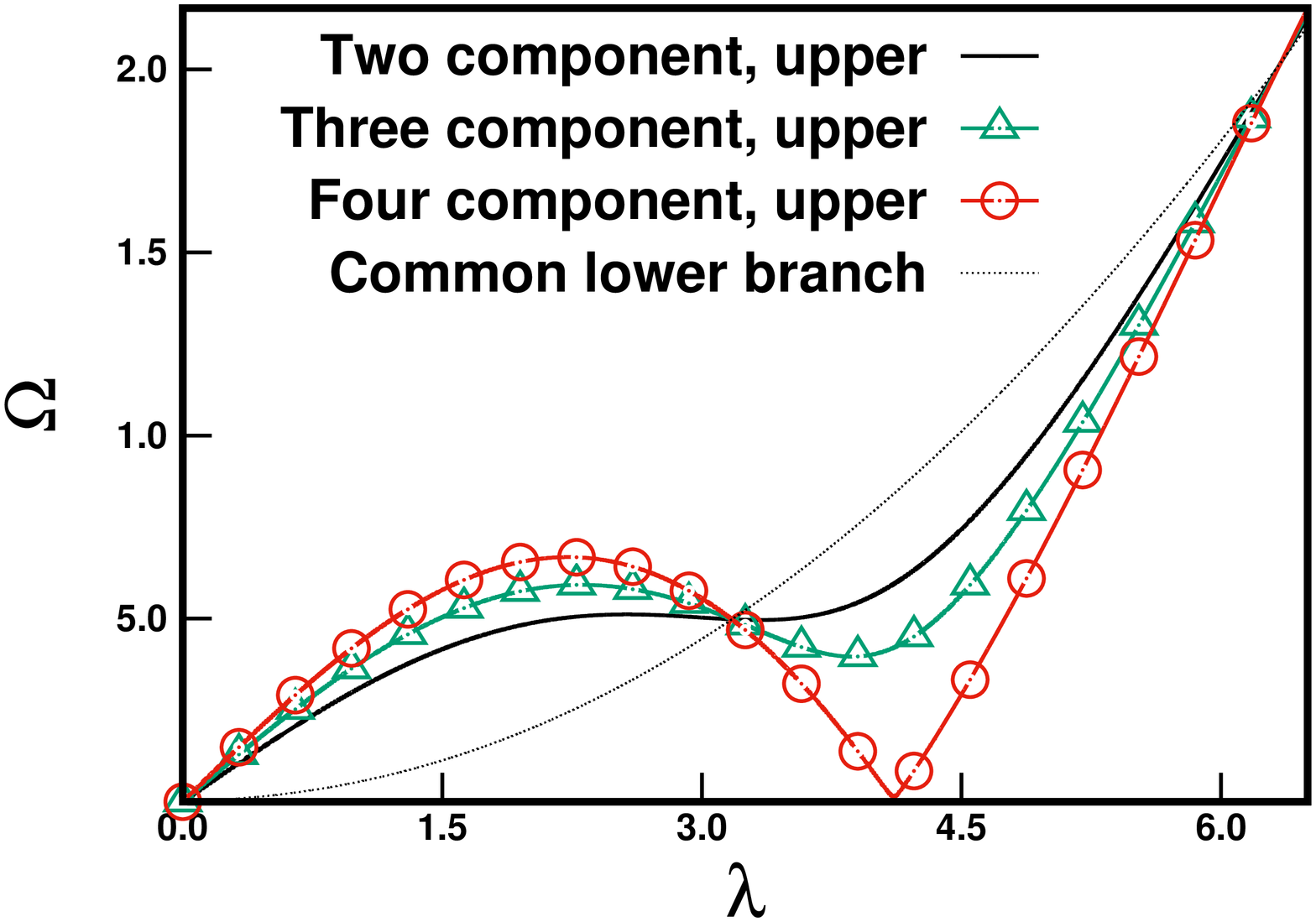}}
        \caption{$\beta=0.99$, $a_{1}^{3}n=0.43$}
        \label{fig:multi_13}
    \end{subfigure}
\hspace*{\fill}

   \begin{subfigure}[b]{0.2\textwidth}
\rotatebox{270}{        \includegraphics[width=\textwidth]{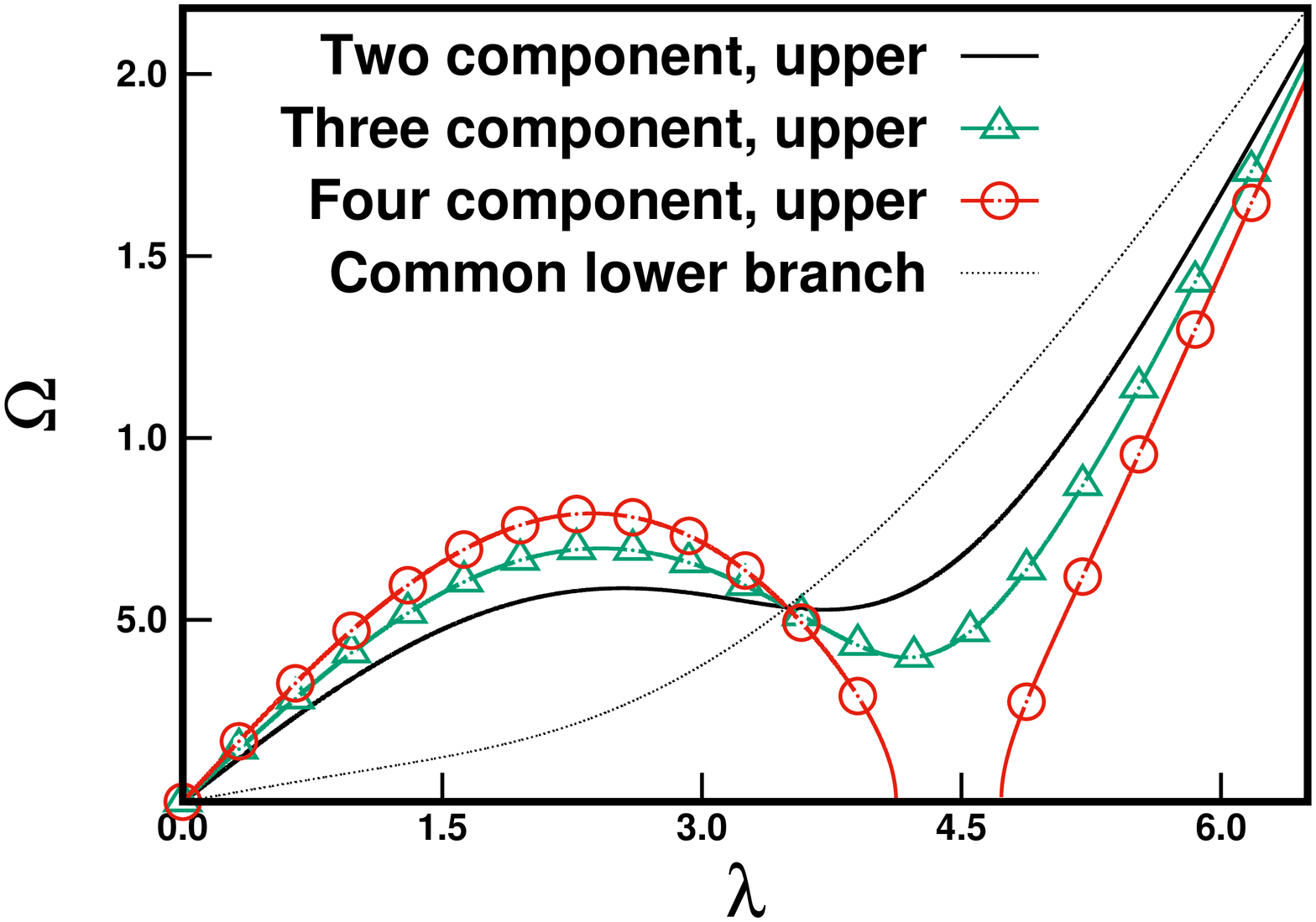}}      
   \caption{$\beta=0.90$, $a_{1}^{3}n=0.57$}
       \label{fig:multi_21}
    \end{subfigure}
\hfill%
    ~ 
   \begin{subfigure}[b]{0.2\textwidth}
\rotatebox{270}{        \includegraphics[width=\textwidth]{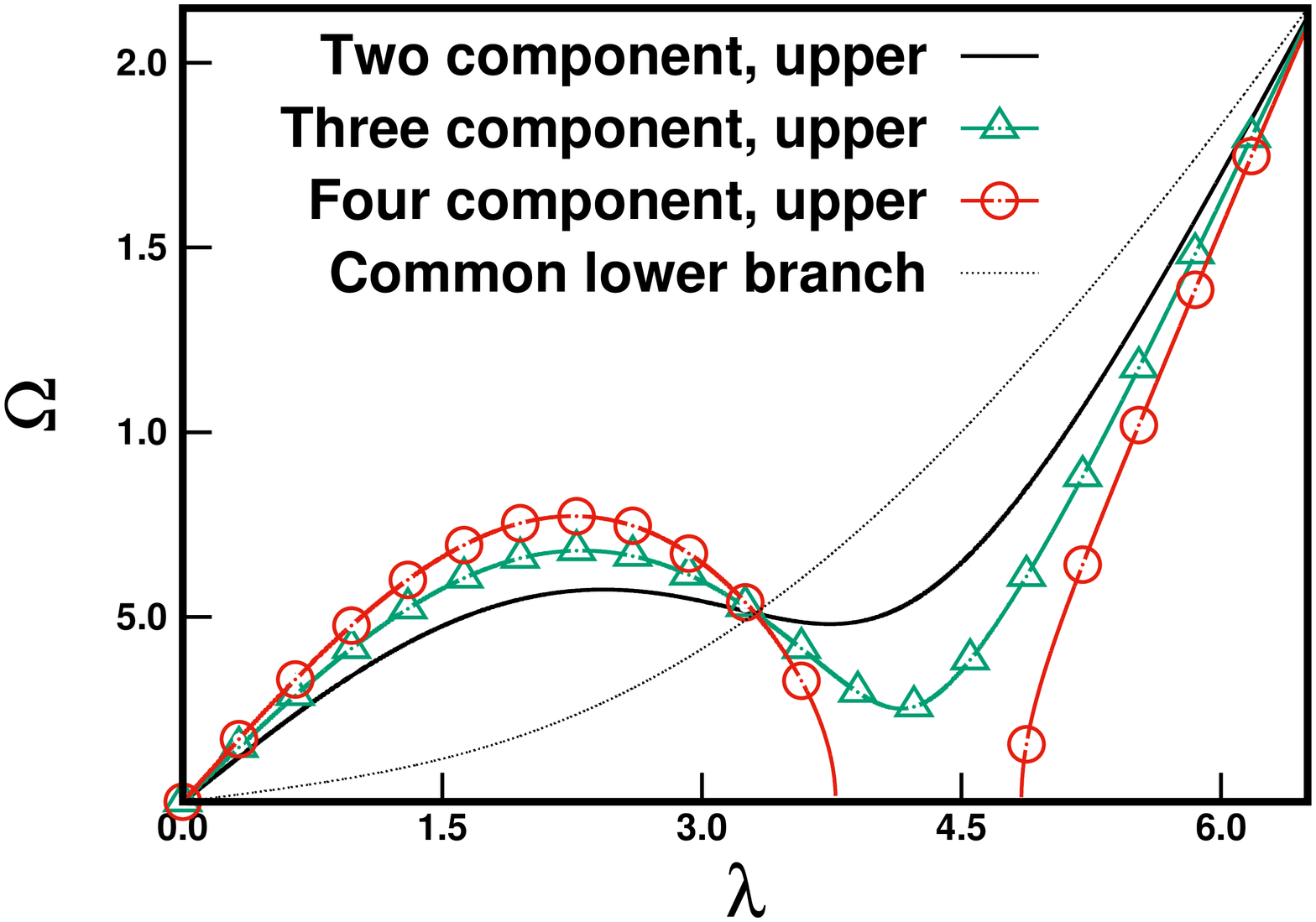}}
        \caption{$\beta=0.95$, $a_{1}^{3}n=0.57$}
        \label{fig:multi_22}
    \end{subfigure}
\hfill
 ~ 
 \begin{subfigure}[b]{0.2\textwidth}
\rotatebox{270}       { \includegraphics[width=\textwidth]{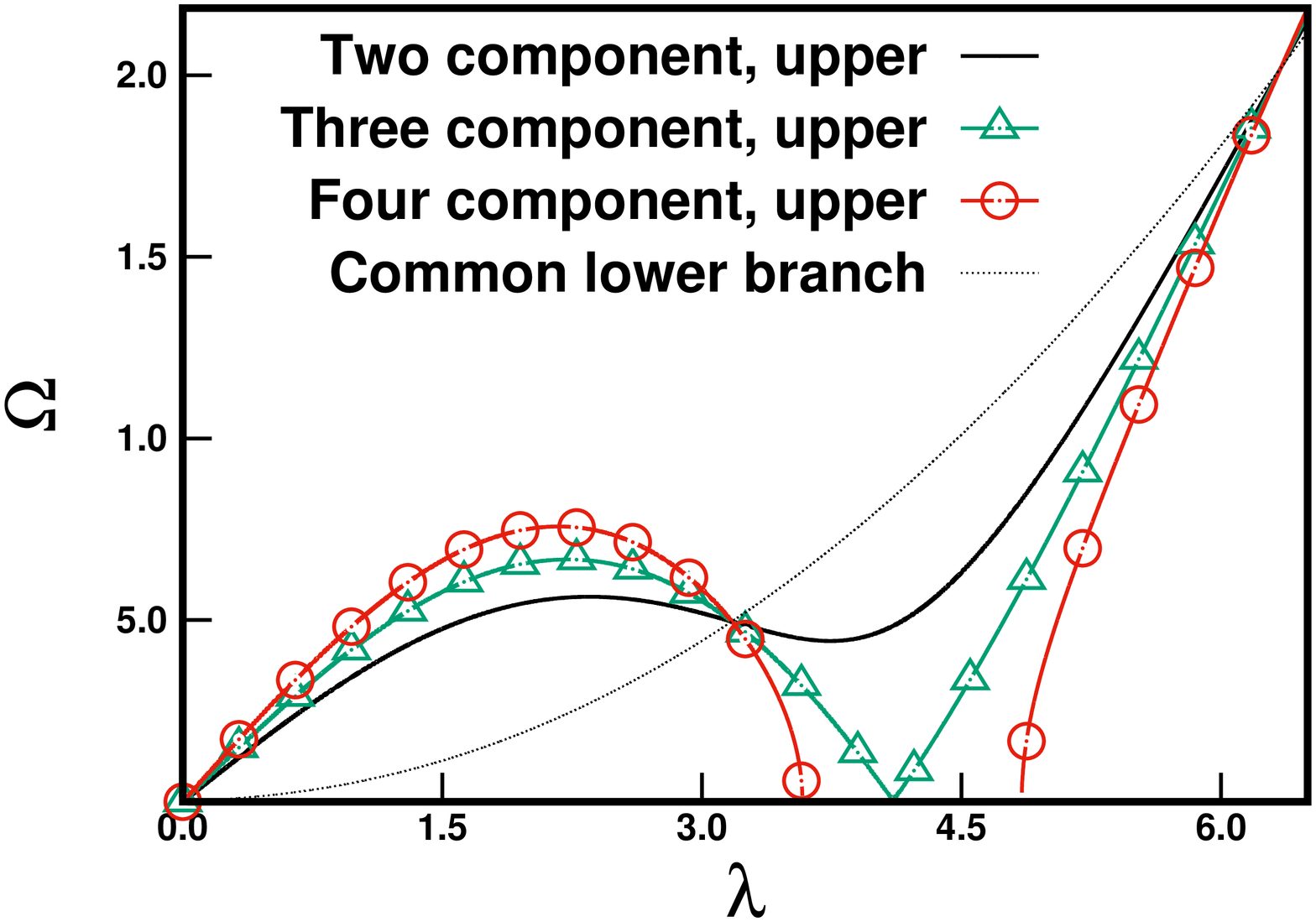}}
        \caption{$\beta=0.99$, $a_{1}^{3}n=0.57$}
        \label{fig:multi_23}
    \end{subfigure}
\hspace*{\fill}%

    \caption{Figures shows the dispersion relation for small amplitude excitations of a BEC, in the presence of non-local interaction pseudopotential between atoms, with various number of distinct components as labelled in the graphs. The plots are for different values of the intra-species gas parameter $a_{1}^{3}n$ and $\beta=a_{2}/a_{1}$.  Here, $\Omega=m\omega a_{1}^{2}/\hbar$ and $\sigma=1$ for the sake of comparison.}
\label{fig:multi}
\end{figure*}

\par
We write the dispersion relation in Eq.(\ref{eq:multi_roton}) in a form similar to that of Eq.(\ref{eq:single_component_roton}) as

\begin{widetext}
\begin{equation}
\begin{split}
\omega=&\frac{\hbar}{a_{1}^{2}m}\Big[\frac{\lambda^{4}}{4}+4\pi a_{1}^{3}n\lambda\sin{\sigma\lambda}+(s-1)4\pi a_{1}^{3}n\lambda\sin{\sigma\beta\lambda}\Big]^{\frac{1}{2}}\\
&\\
& or/and\\
&\\
\omega=&\frac{\hbar}{a_{1}^{2}m}\Big[\frac{\lambda^{4}}{4}+4\pi a_{1}^{3}n\lambda\sin{\sigma\lambda}-4\pi a_{1}^{3}n\lambda\sin{\sigma\beta\lambda}\Big]^{\frac{1}{2}}
\end{split}
\label{eq:multi_roton_scaled}
\end{equation}
\end{widetext}

where $\lambda=a_{1}k$ and $\beta=a_{2}/a_{1}$. As the miscibility criteria states, we should stay in the limit $a_{2}<a_{1}$, implying $\beta<1$ \cite{miscibility_ao}. Also, as we have taken the densities of each component to be the same, the gas parameter takes the form $n a_{2}^{3}$ and $n a_{1}^{3}$ for the inter-species and intra-species scattering length respectively \cite{diluteness_yukalov}. Since $a_{2}<a_{1}$, $na_{2}^{3}<na_{1}^{3}$.

\par
As we had seen in the single component case with non-local interactions, roton lowering was possible due to the $\sin{ak}$ term, since it attains negative values. Here too, the roton minimum is obtained due to $\sin{a_{1}k}$ and $\sin{\beta a_{1}k}$ terms. However, the behaviour of the upper and lower branches of Eq.(\ref{eq:multi_roton_scaled}) is different due to the opposite sign of $\sin{a_{1}k}$ and $\sin{\beta a_{1}k}$, and the absence of $s$ in the lower branch. Also, notice that the lower branch is common to all the multi-component BEC with $s\geq 2$. In Fig.(\ref{fig:lower}) we plot this common branch for various values of the gas parameter. The lowest value of the gas parameter for which the roton mode of this branch may be pulled down to the $\omega=0$ axis is $a_{1}^{3}n\sim 0.616$. Notice here that since $\sin{\lambda}$ and $\sin{\beta \lambda}$ have opposite sign, the roton minimum would be lowered when $\sin{\beta \lambda}$ is positive and $\sin{\lambda}$ is negative. This is in contrast to the upper branch both the terms being negative will enhance the roton lowering. Hence, for the common branch, the value of $\beta=a_{2}/a_{1}$ for which the roton mode softening is enhanced is $\beta\sim 0.49$. For the upper branch, as we shall see, the roton lowering would be enhanced for $\beta$ very close to $1$. Having discussed the common branch, we next discuss the behaviour of the upper branch of Eq.(\ref{eq:multi_roton_scaled}).


\par
In Fig.(\ref{fig:multi}) we plot the dispersion relation of elementary excitations for a two, three and four component BEC for the upper branch in Eq.(\ref{eq:multi_roton_scaled}). This figure shows that, for a fixed value of the parameter $a_{1}^{3}n$, the roton minimum is lowered as we increase the number of distinct components in a BEC. From the figure we also see that the lower branch of $\omega$ in Eq.(\ref{eq:multi_roton_scaled}) does not show any instability in the regime of parameters considered. The values taken in the figure are such that $a_{2}<a_{1}$ such that the miscibility limit is satisfied. For the single component BEC (Fig.(\ref{fig:single})) the roton lowering was achieved at $a^{3}n\sim 1.5$. However, here the roton lowering is achieved at a value of $a_{1}^{3}n\sim a_{2}^{3}n\sim0.43$ for a four component BEC and $a_{1}^{3}n\sim a_{2}^{3}n\sim0.57$ for a three component BEC. Upon further varying the value of $a_{1}^{3}n$ we may observe roton lowering for a two-component BEC in the upper branch. The value of gas parameter for this to happen is $a_{1}^{3}n\sim 0.85$. However, we have already seen that the common branch of excitation for multi-component BEC gives us roton mode softening at a lower value of gas parameter, viz., $a_{1}^{3}n\sim 0.616$. Since, this branch is present for $s\geq 2$, it is present for a two-component BEC as well. Hence the common(lower)branch would give roton lowering for a two component BEC at a smaller value of the gas parameter than the upper branch.

\par
The condition for three body losses to be negligible in a BEC is given by the diluteness condition $a^{3}n<<1$. As the parameter $a^{3}n\rightarrow 1$ or beyond one, loss of atoms from the condensate occurs. Nonetheless, BEC in such strongly interacting regimes have been obtained and studied \cite{tunability_strongly_interacting_pollack}. This is achieved by tuning the s-wave scattering length \cite{scattering_length_chin}. The lifetime of such a strongly interacting BEC would depend on the gas parameter and hence it is important to obtain the roton lowering at a lower value of the gas parameter.

\par
 The loss rate for a BEC is given by the equation $dN(t)/dt=-K_{3}^{tot}<n^{2}>N(t)/6$, where $N(t)$ is the number of atoms at time $t$, $<n^{2}>$ is the mean square density of the condensate and $K_{3}^{tot}$ is the total 3 body loss rate constant\cite{loss_rubidium_smirne}. The factor $K_{3}^{tot}$ is dependent on the gas parameter $a^{3}n$. From Feshbach resonance experiments for commonly used atomic BEC components like rubidium, caesium, potassium, lithium and sodium \cite{loss_rubidium_smirne, loss_caesium_haller, loss_potassium_zaccanti, loss_lithium_gross, loss_sodium_stenger}, one can see that obtaining values of $a^{3}n$ near $1$ requires working very close to the resonance. These experiments have also measured value of $K_{3}^{tot}$ for a range of values of the scattering lengths. One can see from these experiments that $K_{3}^{tot}$ increases very sharply near the resonance. Hence, an increase in the value of $a^{3}n$ from $0.6$ to $1.5$ would increase the loss rate $K_{3}^{tot}$ by two to three orders of magnitude. Thus, it can be seen that the reduction in the value of $a^{3}n$ for roton mode softening that we have shown is quite significant.


\par
We can see that for finite $k$, setting the width of the pseudopotential interaction tending to zero, as is done for local interactions, would mean setting the scattering length tending to zero. One can set $a\rightarrow 0$ in  Eq.(\ref{eq:multi_roton}). This would essentially give us back the multi-component dispersion relation in Eq.(\ref{eq:excitations_local}), which we had got by neglecting the non-locality of the pseudopotential in the multi-component BEC. 
\par
The phenomenon seen in Fig.(\ref{fig:multi}) of the roton mode softening at a lower value of gas parameter, arises mainly because the inter-species interaction. The lower value of gas parameter ensures that the three body losses reduce as we increase the number of components. As the strongly interacting regime $a^{3}n\rightarrow 1$ has been achieved experimentally, such a roton lowering is realizable for a multi-component BEC by tuning the scattering lengths involved at lower loss rates.

\section{Discussion}

\par
In this paper, we have studied elementary excitations in a BEC in the presence of non-local s-wave interactions for multi-component BECs. We started by studying the effect of non-local interactions on the spectrum of elementary excitations in a single component BEC. We considered these elementary excitations on top of the uniform density ground state. The consideration of finite range of inter-boson interactions gives us a roton mode in the spectrum of elementary excitations. This roton mode can be lowered by increasing the gas parameter($a^{3}n$). The roton energy becomes zero for a value of $a^{3}n>1$ for a single component BEC. We then have studied elementary excitations on top of a ground state of multi-component BEC, where each of the components has a uniform density. We work in the limit where the various components are miscible. Consideration of finite range of inter-boson interactions gives us a roton mode for multi-component BEC as well. We investigated to find the lowest value of the gas parameter $a_{1}^{3}n$ for which we get zero energy roton modes. We have shown that for BEC with two components and above, zero energy roton modes can be obtained for $a_{1}^{3}n<1$. Furthermore, we have shown that as we go on increasing the disticnt components of a multi-component BEC, zero energy roton modes are obtained for lower and lower values of $a_{1}^{3}n$.

\par
 Note that we have just considered the s-wave scattering and not any other form of long range interactions. The non-local s-wave scattering pseudopotential is taken to be a rectangular barrier with width of the order of the s-wave scattering length. Previous works have shown that in the presence of non-local interaction pseudopotential, there appears a roton minimum in the dispersion relation, for a single component BEC. This roton minimum can be lowered by modifying the width of interactions, which can be taken to be of the order of the scattering length. However, for achieving this roton mode softening, one has to go beyond the diluteness limit($a^{3}n<<1$) and into the strongly interacting regime.

\par
The experimental progress in obtaining strongly interacting BEC is substantial with the gas parameter being pushed very close to $a^{3}n\sim 1$. This is achieved by tuning the s-wave scattering length. However, as we go near the strongly interacting BEC regime $a^{3}n\sim 1$ value, three body losses become significant. The magnitude of such losses would increase with the increase of the gas parameter. Hence it is necessary to obtain zero energy roton modes at a lower value of the gas parameter and this is where multi-component BEC helps. We have shown that even using just a two-component BEC, we are able to push the value of gas parameter for obtaining zero energy roton modes from $1.5$ to $0.616$. As we increase the number of components further, we get zero energy roton modes for even lower values of the gas parameters.

\par
As the roton mode exhibits itself in a single-component BEC, it is but natural to expect such a roton minimum in a multi-component BEC as well. What is peculiar to the multi-component BEC is the softening of roton mode, at a certain value of gas parameter, just by increasing the number of distinct components. This roton mode softening comes about mainly because of the fact that every species in a multi-component BEC is represented by a particular wave-function. This brings about presence of cross coupling terms in the GP equation used to describe the multi-component system. Due to such terms, the small amplitude oscillation modes of one component couple to the ground state of other component at the expense of energy which helps lower the roton energy. This is seen as roton mode softening in the dispersion relation. Since the strongly interacting regime of a BEC is achieved experimentally, the technique suggested by us would help realize zero energy roton modes at a lower value of gas parameter, thus reducing losses.



\begin{acknowledgments}
I would like to thank my thesis supervisor Dr. Arijit Bhattacharyay for the discussions and help provided to give structure to this work. Also, the input provided by Dr. Peter Mason of Loughborough University was very helpful. This research is funded by the Council of Scientific and Industrial Research(CSIR), India. 
\end{acknowledgments}

\bibliography {references.bib}

\def\germ{\frak} \def\scr{\cal} \ifx\documentclass\undefinedcs
  \def\bf{\fam\bffam\tenbf}\def\rm{\fam0\tenrm}\fi 
  \def\defaultdefine#1#2{\expandafter\ifx\csname#1\endcsname\relax
  \expandafter\def\csname#1\endcsname{#2}\fi} \defaultdefine{Bbb}{\bf}
  \defaultdefine{frak}{\bf} \defaultdefine{=}{\B} 
  \defaultdefine{mathfrak}{\frak} \defaultdefine{mathbb}{\bf}
  \defaultdefine{mathcal}{\cal}
  \defaultdefine{beth}{BETH}\defaultdefine{cal}{\bf} \def\bbfI{{\Bbb I}}
  \def\mbox{\hbox} \def\text{\hbox} \def\om{\omega} \def\Cal#1{{\bf #1}}
  \def\pcf{pcf} \defaultdefine{cf}{cf} \defaultdefine{reals}{{\Bbb R}}
  \defaultdefine{real}{{\Bbb R}} \def\restriction{{|}} \def\club{CLUB}
  \def\w{\omega} \def\exist{\exists} \def\se{{\germ se}} \def\bb{{\bf b}}
  \def\equivalence{\equiv} \let\lt< \let\gt>
\begin{thebibliography}{10}%
\makeatletter
\providecommand \@ifxundefined [1]{%
 \ifx #1\undefined \expandafter \@firstoftwo
 \else \expandafter \@secondoftwo
\fi
}%
\providecommand \@ifnum [1]{%
 \ifnum #1\expandafter \@firstoftwo
 \else \expandafter \@secondoftwo
\fi
}%
\providecommand \enquote [1]{``#1''}%
\providecommand \bibnamefont  [1]{#1}%
\providecommand \bibfnamefont [1]{#1}%
\providecommand \citenamefont [1]{#1}%
\providecommand\href[0]{\@sanitize\@href}%
\providecommand\@href[1]{\endgroup\@@startlink{#1}\endgroup\@@href}%
\providecommand\@@href[1]{#1\@@endlink}%
\providecommand \@sanitize [0]{\begingroup\catcode`\&12\catcode`\#12\relax}%
\@ifxundefined \pdfoutput {\@firstoftwo}{%
 \@ifnum{\z@=\pdfoutput}{\@firstoftwo}{\@secondoftwo}%
}{%
 \providecommand\@@startlink[1]{\leavevmode\special{html:<a href="#1">}}%
 \providecommand\@@endlink[0]{\special{html:</a>}}%
}{%
 \providecommand\@@startlink[1]{%
  \leavevmode
  \pdfstartlink
   attr{/Border[0 0 1 ]/H/I/C[0 1 1]}%
   user{/Subtype/Link/A<</Type/Action/S/URI/URI(#1)>>}%
  \relax
 }%
 \providecommand\@@endlink[0]{\pdfendlink}%
}%
\providecommand \url  [0]{\begingroup\@sanitize \@url }%
\providecommand \@url [1]{\endgroup\@href {#1}{\urlprefix}}%
\providecommand \urlprefix [0]{URL }%
\providecommand \Eprint[0]{\href }%
\@ifxundefined \urlstyle {%
  \providecommand \doi [1]{doi:\discretionary{}{}{}#1}%
}{%
  \providecommand \doi [0]{doi:\discretionary{}{}{}\begingroup
  \urlstyle{rm}\Url }%
}%
\providecommand \doibase [0]{http://dx.doi.org/}%
\providecommand \Doi[1]{\href{\doibase#1}}%
\providecommand \bibAnnote [3]{%
  \BibitemShut{#1}%
  \begin{quotation}\noindent
    \textsc{Key:}\ #2\\\textsc{Annotation:}\ #3%
  \end{quotation}%
}%
\providecommand \bibAnnoteFile [2]{%
  \IfFileExists{#2}{\bibAnnote {#1} {#2} {\input{#2}}}{}%
}%
\providecommand \typeout [0]{\immediate \write \m@ne }%
\providecommand \selectlanguage [0]{\@gobble}%
\providecommand \bibinfo [0]{\@secondoftwo}%
\providecommand \bibfield [0]{\@secondoftwo}%
\providecommand \translation [1]{[#1]}%
\providecommand \BibitemOpen[0]{}%
\providecommand \bibitemStop [0]{}%
\providecommand \bibitemNoStop [0]{.\EOS\space}%
\providecommand \EOS [0]{\spacefactor3000\relax}%
\providecommand \BibitemShut [1]{\csname bibitem#1\endcsname}%
\bibitem{atomicbec_anderson}%
  \BibitemOpen
  \bibfield{author}{%
  \bibinfo {author} {\bibfnamefont{M.~H.}\ \bibnamefont{Anderson}}, \bibinfo
  {author} {\bibfnamefont{J.~R.}\ \bibnamefont{Ensher}}, \bibinfo {author}
  {\bibfnamefont{M.~R.}\ \bibnamefont{Matthews}}, \bibinfo {author}
  {\bibfnamefont{C.~E.}\ \bibnamefont{Wieman}},\ and\ \bibinfo {author}
  {\bibfnamefont{E.~A.}\ \bibnamefont{Cornell}},\ }%
  \bibfield{journal}{%
  \bibinfo {journal} {Science}\ }%
  \textbf{\bibinfo {volume} {269}},\ \bibinfo {pages} {198} (\bibinfo {year}
  {1995})%
  \bibAnnoteFile{NoStop}{atomicbec_anderson}%
\bibitem{atomicbec_ketterle}%
  \BibitemOpen
  \bibfield{author}{%
  \bibinfo {author} {\bibfnamefont{W.~K.}\ \bibnamefont{et~al.}},\ }%
  \bibfield{journal}{%
  \bibinfo {journal} {Phys. Rev. Lett.}\ }%
  \textbf{\bibinfo {volume} {75, 3969}} (\bibinfo {year} {1995})%
  \bibAnnoteFile{NoStop}{atomicbec_ketterle}%
\bibitem{ps}%
  \BibitemOpen
  \bibfield{author}{%
  \bibinfo {author} {\bibfnamefont{L.}~\bibnamefont{Pitaevskii}}\ and\ \bibinfo
  {author} {\bibfnamefont{S.}~\bibnamefont{Stringari}},\ }%
  \emph{\bibinfo {title} {Bose-Einstein Condensation}}\ (\bibinfo {publisher}
  {Oxford Science Publications},\ \bibinfo {year} {2003})%
  \bibAnnoteFile{NoStop}{ps}%
\bibitem{superfluidity_landau}%
  \BibitemOpen
  \bibfield{author}{%
  \bibinfo {author} {\bibfnamefont{L.}~\bibnamefont{Landau}},\ }%
  \bibfield{journal}{%
  \bibinfo {journal} {Phys. Rev.}\ }%
  \textbf{\bibinfo {volume} {75}} (\bibinfo {year} {1949})%
  \bibAnnoteFile{NoStop}{superfluidity_landau}%
\bibitem{supersolid_leggett}%
  \BibitemOpen
  \bibfield{author}{%
  \bibinfo {author} {\bibfnamefont{A.~J.}\ \bibnamefont{Leggett}},\ }%
  \bibfield{journal}{%
  \bibinfo {journal} {Phys. Rev. Lett.}\ }%
  \textbf{\bibinfo {volume} {25}},\ \bibinfo {pages} {1543} (\bibinfo {year}
  {1970})%
  \bibAnnoteFile{NoStop}{supersolid_leggett}%
\bibitem{quantumcrystals_chester}%
  \BibitemOpen
  \bibfield{author}{%
  \bibinfo {author} {\bibfnamefont{G.}~\bibnamefont{Chester}},\ }%
  \bibfield{journal}{%
  \bibinfo {journal} {Phys. Rev. A}\ }%
  \textbf{\bibinfo {volume} {2}},\ \bibinfo {pages} {256} (\bibinfo {year}
  {1970})%
  \bibAnnoteFile{NoStop}{quantumcrystals_chester}%
\bibitem{roton_rydberg_henkel}%
  \BibitemOpen
  \bibfield{author}{%
  \bibinfo {author} {\bibfnamefont{N.}~\bibnamefont{Henkel}}, \bibinfo {author}
  {\bibfnamefont{R.}~\bibnamefont{Nath}},\ and\ \bibinfo {author}
  {\bibfnamefont{T.}~\bibnamefont{Pohl}},\ }%
  \bibfield{journal}{%
  \bibinfo {journal} {Physical review letters}\ }%
  \textbf{\bibinfo {volume} {104}},\ \bibinfo {pages} {195302} (\bibinfo {year}
  {2010})%
  \bibAnnoteFile{NoStop}{roton_rydberg_henkel}%
\bibitem{roton_softcore_ancilotto}%
  \BibitemOpen
  \bibfield{author}{%
  \bibinfo {author} {\bibfnamefont{F.}~\bibnamefont{Ancilotto}}, \bibinfo
  {author} {\bibfnamefont{M.}~\bibnamefont{Rossi}},\ and\ \bibinfo {author}
  {\bibfnamefont{F.}~\bibnamefont{Toigo}},\ }%
  \bibfield{journal}{%
  \bibinfo {journal} {Physical Review A}\ }%
  \textbf{\bibinfo {volume} {88}},\ \bibinfo {pages} {033618} (\bibinfo {year}
  {2013})%
  \bibAnnoteFile{NoStop}{roton_softcore_ancilotto}%
\bibitem{roton_experiment_mottl}%
  \BibitemOpen
  \bibfield{author}{%
  \bibinfo {author} {\bibfnamefont{R.}~\bibnamefont{Mottl}}, \bibinfo {author}
  {\bibfnamefont{F.}~\bibnamefont{Brennecke}}, \bibinfo {author}
  {\bibfnamefont{K.}~\bibnamefont{Baumann}}, \bibinfo {author}
  {\bibfnamefont{R.}~\bibnamefont{Landig}}, \bibinfo {author}
  {\bibfnamefont{T.}~\bibnamefont{Donner}},\ and\ \bibinfo {author}
  {\bibfnamefont{T.}~\bibnamefont{Esslinger}},\ }%
  \bibfield{journal}{%
  \bibinfo {journal} {Science}\ }%
  \textbf{\bibinfo {volume} {336}},\ \bibinfo {pages} {1570} (\bibinfo {year}
  {2012})%
  \bibAnnoteFile{NoStop}{roton_experiment_mottl}%
\bibitem{roton_blakie}%
  \BibitemOpen
  \bibfield{author}{%
  \bibinfo {author} {\bibfnamefont{P.}~\bibnamefont{Blakie}}, \bibinfo {author}
  {\bibfnamefont{D.}~\bibnamefont{Baillie}},\ and\ \bibinfo {author}
  {\bibfnamefont{R.}~\bibnamefont{Bisset}},\ }%
  \bibfield{journal}{%
  \bibinfo {journal} {Physical Review A}\ }%
  \textbf{\bibinfo {volume} {86}},\ \bibinfo {pages} {021604} (\bibinfo {year}
  {2012})%
  \bibAnnoteFile{NoStop}{roton_blakie}%
\bibitem{roton_santos}%
  \BibitemOpen
  \bibfield{author}{%
  \bibinfo {author} {\bibfnamefont{L.}~\bibnamefont{Santos}}, \bibinfo {author}
  {\bibfnamefont{G.}~\bibnamefont{Shlyapnikov}},\ and\ \bibinfo {author}
  {\bibfnamefont{M.}~\bibnamefont{Lewenstein}},\ }%
  \bibfield{journal}{%
  \bibinfo {journal} {Physical review letters}\ }%
  \textbf{\bibinfo {volume} {90}},\ \bibinfo {pages} {250403} (\bibinfo {year}
  {2003})%
  \bibAnnoteFile{NoStop}{roton_santos}%
\bibitem{nonlocal_pomeau}%
  \BibitemOpen
  \bibfield{author}{%
  \bibinfo {author} {\bibfnamefont{Y.}~\bibnamefont{Pomeau}}\ and\ \bibinfo
  {author} {\bibfnamefont{S.}~\bibnamefont{Rica}},\ }%
  \bibfield{journal}{%
  \bibinfo {journal} {Physical review letters}\ }%
  \textbf{\bibinfo {volume} {72}},\ \bibinfo {pages} {2426} (\bibinfo {year}
  {1994})%
  \bibAnnoteFile{NoStop}{nonlocal_pomeau}%
\bibitem{nonlocal_pseudopotential_macri}%
  \BibitemOpen
  \bibfield{author}{%
  \bibinfo {author} {\bibfnamefont{T.}~\bibnamefont{Macri}}, \bibinfo {author}
  {\bibfnamefont{F.}~\bibnamefont{Maucher}}, \bibinfo {author}
  {\bibfnamefont{F.}~\bibnamefont{Cinti}},\ and\ \bibinfo {author}
  {\bibfnamefont{T.}~\bibnamefont{Pohl}},\ }%
  \bibfield{journal}{%
  \bibinfo {journal} {Physical Review A}\ }%
  \textbf{\bibinfo {volume} {87}},\ \bibinfo {pages} {061602} (\bibinfo {year}
  {2013})%
  \bibAnnoteFile{NoStop}{nonlocal_pseudopotential_macri}%
\bibitem{feshbach_inouye}%
  \BibitemOpen
  \bibfield{author}{%
  \bibinfo {author} {\bibfnamefont{S.}~\bibnamefont{Inouye}}, \bibinfo {author}
  {\bibfnamefont{M.}~\bibnamefont{Andrews}}, \bibinfo {author}
  {\bibfnamefont{J.}~\bibnamefont{Stenger}}, \bibinfo {author}
  {\bibfnamefont{H.-J.}\ \bibnamefont{Miesner}}, \bibinfo {author}
  {\bibfnamefont{D.}~\bibnamefont{Stamper-Kurn}},\ and\ \bibinfo {author}
  {\bibfnamefont{W.}~\bibnamefont{Ketterle}},\ }%
  \bibfield{journal}{%
  \bibinfo {journal} {Nature}\ }%
  \textbf{\bibinfo {volume} {392}},\ \bibinfo {pages} {151} (\bibinfo {year}
  {1998})%
  \bibAnnoteFile{NoStop}{feshbach_inouye}%
\bibitem{twocomponent_cornish}%
  \BibitemOpen
  \bibfield{author}{%
  \bibinfo {author} {\bibfnamefont{D.}~\bibnamefont{McCarron}}, \bibinfo
  {author} {\bibfnamefont{H.}~\bibnamefont{Cho}}, \bibinfo {author}
  {\bibfnamefont{D.}~\bibnamefont{Jenkin}}, \bibinfo {author}
  {\bibfnamefont{M.}~\bibnamefont{K{\"o}ppinger}},\ and\ \bibinfo {author}
  {\bibfnamefont{S.}~\bibnamefont{Cornish}},\ }%
  \bibfield{journal}{%
  \bibinfo {journal} {Physical Review A}\ }%
  \textbf{\bibinfo {volume} {84}},\ \bibinfo {pages} {011603} (\bibinfo {year}
  {2011})%
  \bibAnnoteFile{NoStop}{twocomponent_cornish}%
\bibitem{yb_yoshiro}%
  \BibitemOpen
  \bibfield{author}{%
  \bibinfo {author} {\bibfnamefont{T.}~\bibnamefont{Fukuhara}}, \bibinfo
  {author} {\bibfnamefont{S.}~\bibnamefont{Sugawa}},\ and\ \bibinfo {author}
  {\bibfnamefont{Y.}~\bibnamefont{Takahashi}},\ }%
  \bibfield{journal}{%
  \bibinfo {journal} {Physical Review A}\ }%
  \textbf{\bibinfo {volume} {76}},\ \bibinfo {pages} {051604} (\bibinfo {year}
  {2007})%
  \bibAnnoteFile{NoStop}{yb_yoshiro}%
\bibitem{multicomponent_ueda}%
  \BibitemOpen
  \bibfield{author}{%
  \bibinfo {author} {\bibfnamefont{D.~C.}\ \bibnamefont{Roberts}}\ and\
  \bibinfo {author} {\bibfnamefont{M.}~\bibnamefont{Ueda}},\ }%
  \bibfield{journal}{%
  \bibinfo {journal} {Physical Review A}\ }%
  \textbf{\bibinfo {volume} {73}},\ \bibinfo {pages} {053611} (\bibinfo {year}
  {2006})%
  \bibAnnoteFile{NoStop}{multicomponent_ueda}%
\bibitem{miscibility_twocomponent_wieman}%
  \BibitemOpen
  \bibfield{author}{%
  \bibinfo {author} {\bibfnamefont{S.}~\bibnamefont{Papp}}, \bibinfo {author}
  {\bibfnamefont{J.}~\bibnamefont{Pino}},\ and\ \bibinfo {author}
  {\bibfnamefont{C.}~\bibnamefont{Wieman}},\ }%
  \bibfield{journal}{%
  \bibinfo {journal} {Physical review letters}\ }%
  \textbf{\bibinfo {volume} {101}},\ \bibinfo {pages} {040402} (\bibinfo {year}
  {2008})%
  \bibAnnoteFile{NoStop}{miscibility_twocomponent_wieman}%
\bibitem{stability_twocomponent_savage}%
  \BibitemOpen
  \bibfield{author}{%
  \bibinfo {author} {\bibfnamefont{D.}~\bibnamefont{Gordon}}\ and\ \bibinfo
  {author} {\bibfnamefont{C.}~\bibnamefont{Savage}},\ }%
  \bibfield{journal}{%
  \bibinfo {journal} {Physical Review A}\ }%
  \textbf{\bibinfo {volume} {58}},\ \bibinfo {pages} {1440} (\bibinfo {year}
  {1998})%
  \bibAnnoteFile{NoStop}{stability_twocomponent_savage}%
\bibitem{interspecies_interaction_inguscio}%
  \BibitemOpen
  \bibfield{author}{%
  \bibinfo {author} {\bibfnamefont{G.}~\bibnamefont{Thalhammer}}, \bibinfo
  {author} {\bibfnamefont{G.}~\bibnamefont{Barontini}}, \bibinfo {author}
  {\bibfnamefont{L.}~\bibnamefont{De~Sarlo}}, \bibinfo {author}
  {\bibfnamefont{J.}~\bibnamefont{Catani}}, \bibinfo {author}
  {\bibfnamefont{F.}~\bibnamefont{Minardi}},\ and\ \bibinfo {author}
  {\bibfnamefont{M.}~\bibnamefont{Inguscio}},\ }%
  \bibfield{journal}{%
  \bibinfo {journal} {Physical review letters}\ }%
  \textbf{\bibinfo {volume} {100}},\ \bibinfo {pages} {210402} (\bibinfo {year}
  {2008})%
  \bibAnnoteFile{NoStop}{interspecies_interaction_inguscio}%
\bibitem{gp_gross}%
  \BibitemOpen
  \bibfield{author}{%
  \bibinfo {author} {\bibfnamefont{E.~P.}\ \bibnamefont{Gross}},\ }%
  \bibfield{journal}{%
  \bibinfo {journal} {Annals of Physics}\ }%
  \textbf{\bibinfo {volume} {9}},\ \bibinfo {pages} {292} (\bibinfo {year}
  {1960})%
  \bibAnnoteFile{NoStop}{gp_gross}%
\bibitem{pethick}%
  \BibitemOpen
  \bibfield{author}{%
  \bibinfo {author} {\bibfnamefont{C.}~\bibnamefont{Pethick}}\ and\ \bibinfo
  {author} {\bibfnamefont{H.}~\bibnamefont{Smith}},\ }%
  \emph{\bibinfo {title} {Bose-Einstein Condensation in Dilute Gases}}\
  (\bibinfo {publisher} {Cambridge University Press},\ \bibinfo {year} {2001})%
  \bibAnnoteFile{NoStop}{pethick}%
\bibitem{nonlocal_ketterle}%
  \BibitemOpen
  \bibfield{author}{%
  \bibinfo {author} {\bibfnamefont{H.}~\bibnamefont{Veksler}}, \bibinfo
  {author} {\bibfnamefont{S.}~\bibnamefont{Fishman}},\ and\ \bibinfo {author}
  {\bibfnamefont{W.}~\bibnamefont{Ketterle}},\ }%
  \bibfield{journal}{%
  \bibinfo {journal} {Physical Review A}\ }%
  \textbf{\bibinfo {volume} {90}},\ \bibinfo {pages} {023620} (\bibinfo {year}
  {2014})%
  \bibAnnoteFile{NoStop}{nonlocal_ketterle}%
\bibitem{nonlocal_kutz}%
  \BibitemOpen
  \bibfield{author}{%
  \bibinfo {author} {\bibfnamefont{J.~N.}\ \bibnamefont{Kutz}}\ and\ \bibinfo
  {author} {\bibfnamefont{B.}~\bibnamefont{Deconinck}},\ }%
  in\ \emph{\bibinfo {booktitle} {Nonlinear Guided Waves and Their
  Applications}}\ (\bibinfo {organization} {Optical Society of America},\
  \bibinfo {year} {2002})\ p.\ \bibinfo {pages} {NLTuD40}%
  \bibAnnoteFile{NoStop}{nonlocal_kutz}%
\bibitem{nonlocal_pethick}%
  \BibitemOpen
  \bibfield{author}{%
  \bibinfo {author} {\bibfnamefont{A.}~\bibnamefont{Collin}}, \bibinfo {author}
  {\bibfnamefont{P.}~\bibnamefont{Massignan}},\ and\ \bibinfo {author}
  {\bibfnamefont{C.}~\bibnamefont{Pethick}},\ }%
  \bibfield{journal}{%
  \bibinfo {journal} {Physical Review A}\ }%
  \textbf{\bibinfo {volume} {75}},\ \bibinfo {pages} {013615} (\bibinfo {year}
  {2007})%
  \bibAnnoteFile{NoStop}{nonlocal_pethick}%
\bibitem{bragg_strongly_interacting_papp}%
  \BibitemOpen
  \bibfield{author}{%
  \bibinfo {author} {\bibfnamefont{S.}~\bibnamefont{Papp}}, \bibinfo {author}
  {\bibfnamefont{J.}~\bibnamefont{Pino}}, \bibinfo {author}
  {\bibfnamefont{R.}~\bibnamefont{Wild}}, \bibinfo {author}
  {\bibfnamefont{S.}~\bibnamefont{Ronen}}, \bibinfo {author}
  {\bibfnamefont{C.~E.}\ \bibnamefont{Wieman}}, \bibinfo {author}
  {\bibfnamefont{D.~S.}\ \bibnamefont{Jin}},\ and\ \bibinfo {author}
  {\bibfnamefont{E.~A.}\ \bibnamefont{Cornell}},\ }%
  \bibfield{journal}{%
  \bibinfo {journal} {Physical review letters}\ }%
  \textbf{\bibinfo {volume} {101}},\ \bibinfo {pages} {135301} (\bibinfo {year}
  {2008})%
  \bibAnnoteFile{NoStop}{bragg_strongly_interacting_papp}%
\bibitem{tunability_strongly_interacting_pollack}%
  \BibitemOpen
  \bibfield{author}{%
  \bibinfo {author} {\bibfnamefont{S.~E.}\ \bibnamefont{Pollack}}, \bibinfo
  {author} {\bibfnamefont{D.}~\bibnamefont{Dries}}, \bibinfo {author}
  {\bibfnamefont{M.}~\bibnamefont{Junker}}, \bibinfo {author}
  {\bibfnamefont{Y.}~\bibnamefont{Chen}}, \bibinfo {author}
  {\bibfnamefont{T.}~\bibnamefont{Corcovilos}},\ and\ \bibinfo {author}
  {\bibfnamefont{R.}~\bibnamefont{Hulet}},\ }%
  \bibfield{journal}{%
  \bibinfo {journal} {Physical Review Letters}\ }%
  \textbf{\bibinfo {volume} {102}},\ \bibinfo {pages} {090402} (\bibinfo {year}
  {2009})%
  \bibAnnoteFile{NoStop}{tunability_strongly_interacting_pollack}%
\bibitem{diluteness_yukalov}%
  \BibitemOpen
  \bibfield{author}{%
  \bibinfo {author} {\bibfnamefont{V.}~\bibnamefont{Yukalov}}\ and\ \bibinfo
  {author} {\bibfnamefont{E.}~\bibnamefont{Yukalova}},\ }%
  \bibfield{journal}{%
  \bibinfo {journal} {Laser Physics Letters}\ }%
  \textbf{\bibinfo {volume} {1}},\ \bibinfo {pages} {50} (\bibinfo {year}
  {2004})%
  \bibAnnoteFile{NoStop}{diluteness_yukalov}%
\bibitem{sakurai}%
  \BibitemOpen
  \bibfield{author}{%
  \bibinfo {author} {\bibfnamefont{J.~J.}\ \bibnamefont{Sakurai}}\ and\
  \bibinfo {author} {\bibfnamefont{J.}~\bibnamefont{Napolitano}},\ }%
  \emph{\bibinfo {title} {Modern quantum mechanics}}\ (\bibinfo {publisher}
  {Addison-Wesley},\ \bibinfo {year} {2011})%
  \bibAnnoteFile{NoStop}{sakurai}%
\bibitem{thesis_pattinson}%
  \BibitemOpen
  \bibfield{author}{%
  \bibinfo {author} {\bibfnamefont{R.~W.}\ \bibnamefont{Pattinson}},\ }%
  \emph{\bibinfo {title} {Two-component Bose-Einstein condensates: equilibria
  and dynamics at zero temperature and beyond}}\ (\bibinfo {publisher}
  {Newcastle University},\ \bibinfo {year} {2014})%
  \bibAnnoteFile{NoStop}{thesis_pattinson}%
\bibitem{miscibility_ao}%
  \BibitemOpen
  \bibfield{author}{%
  \bibinfo {author} {\bibfnamefont{P.}~\bibnamefont{Ao}}\ and\ \bibinfo
  {author} {\bibfnamefont{S.}~\bibnamefont{Chui}},\ }%
  \bibfield{journal}{%
  \bibinfo {journal} {Physical Review A}\ }%
  \textbf{\bibinfo {volume} {58}},\ \bibinfo {pages} {4836} (\bibinfo {year}
  {1998})%
  \bibAnnoteFile{NoStop}{miscibility_ao}%
\bibitem{two_component_excitations_alexandrov}%
  \BibitemOpen
  \bibfield{author}{%
  \bibinfo {author} {\bibfnamefont{A.}~\bibnamefont{Alexandrov}}\ and\ \bibinfo
  {author} {\bibfnamefont{V.~V.}\ \bibnamefont{Kabanov}},\ }%
  \bibfield{journal}{%
  \bibinfo {journal} {Journal of Physics: Condensed Matter}\ }%
  \textbf{\bibinfo {volume} {14}},\ \bibinfo {pages} {L327} (\bibinfo {year}
  {2002})%
  \bibAnnoteFile{NoStop}{two_component_excitations_alexandrov}%
\bibitem{scattering_length_chin}%
  \BibitemOpen
  \bibfield{author}{%
  \bibinfo {author} {\bibfnamefont{C.}~\bibnamefont{Chin}}\ and\ \bibinfo
  {author} {\bibfnamefont{V.~V.}\ \bibnamefont{Flambaum}},\ }%
  \bibfield{journal}{%
  \bibinfo {journal} {Phys. Rev. Lett.}\ }%
  \textbf{\bibinfo {volume} {96}},\ \bibinfo {pages} {230801} (\bibinfo {year}
  {2006})%
  \bibAnnoteFile{NoStop}{scattering_length_chin}%
\bibitem{loss_rubidium_smirne}%
  \BibitemOpen
  \bibfield{author}{%
  \bibinfo {author} {\bibfnamefont{G.}~\bibnamefont{Smirne}}, \bibinfo {author}
  {\bibfnamefont{R.}~\bibnamefont{Godun}}, \bibinfo {author}
  {\bibfnamefont{D.}~\bibnamefont{Cassettari}}, \bibinfo {author}
  {\bibfnamefont{V.}~\bibnamefont{Boyer}}, \bibinfo {author}
  {\bibfnamefont{C.}~\bibnamefont{Foot}}, \bibinfo {author}
  {\bibfnamefont{T.}~\bibnamefont{Volz}}, \bibinfo {author}
  {\bibfnamefont{N.}~\bibnamefont{Syassen}}, \bibinfo {author}
  {\bibfnamefont{S.}~\bibnamefont{D{\"u}rr}}, \bibinfo {author}
  {\bibfnamefont{G.}~\bibnamefont{Rempe}}, \bibinfo {author}
  {\bibfnamefont{M.}~\bibnamefont{Lee}}, \emph{et~al.},\ }%
  \bibfield{journal}{%
  \bibinfo {journal} {Physical Review A}\ }%
  \textbf{\bibinfo {volume} {75}},\ \bibinfo {pages} {020702} (\bibinfo {year}
  {2007})%
  \bibAnnoteFile{NoStop}{loss_rubidium_smirne}%
\bibitem{loss_caesium_haller}%
  \BibitemOpen
  \bibfield{author}{%
  \bibinfo {author} {\bibfnamefont{E.}~\bibnamefont{Haller}}, \bibinfo {author}
  {\bibfnamefont{M.}~\bibnamefont{Rabie}}, \bibinfo {author}
  {\bibfnamefont{M.~J.}\ \bibnamefont{Mark}}, \bibinfo {author}
  {\bibfnamefont{J.~G.}\ \bibnamefont{Danzl}}, \bibinfo {author}
  {\bibfnamefont{R.}~\bibnamefont{Hart}}, \bibinfo {author}
  {\bibfnamefont{K.}~\bibnamefont{Lauber}}, \bibinfo {author}
  {\bibfnamefont{G.}~\bibnamefont{Pupillo}},\ and\ \bibinfo {author}
  {\bibfnamefont{H.-C.}\ \bibnamefont{N{\"a}gerl}},\ }%
  \bibfield{journal}{%
  \bibinfo {journal} {Physical review letters}\ }%
  \textbf{\bibinfo {volume} {107}},\ \bibinfo {pages} {230404} (\bibinfo {year}
  {2011})%
  \bibAnnoteFile{NoStop}{loss_caesium_haller}%
\bibitem{loss_potassium_zaccanti}%
  \BibitemOpen
  \bibfield{author}{%
  \bibinfo {author} {\bibfnamefont{M.}~\bibnamefont{Zaccanti}}, \bibinfo
  {author} {\bibfnamefont{B.}~\bibnamefont{Deissler}}, \bibinfo {author}
  {\bibfnamefont{C.}~\bibnamefont{D’Errico}}, \bibinfo {author}
  {\bibfnamefont{M.}~\bibnamefont{Fattori}}, \bibinfo {author}
  {\bibfnamefont{M.}~\bibnamefont{Jona-Lasinio}}, \bibinfo {author}
  {\bibfnamefont{S.}~\bibnamefont{M{\"u}ller}}, \bibinfo {author}
  {\bibfnamefont{G.}~\bibnamefont{Roati}}, \bibinfo {author}
  {\bibfnamefont{M.}~\bibnamefont{Inguscio}},\ and\ \bibinfo {author}
  {\bibfnamefont{G.}~\bibnamefont{Modugno}},\ }%
  \bibfield{journal}{%
  \bibinfo {journal} {Nature Physics}\ }%
  \textbf{\bibinfo {volume} {5}},\ \bibinfo {pages} {586} (\bibinfo {year}
  {2009})%
  \bibAnnoteFile{NoStop}{loss_potassium_zaccanti}%
\bibitem{loss_lithium_gross}%
  \BibitemOpen
  \bibfield{author}{%
  \bibinfo {author} {\bibfnamefont{N.}~\bibnamefont{Gross}}, \bibinfo {author}
  {\bibfnamefont{Z.}~\bibnamefont{Shotan}}, \bibinfo {author}
  {\bibfnamefont{S.}~\bibnamefont{Kokkelmans}},\ and\ \bibinfo {author}
  {\bibfnamefont{L.}~\bibnamefont{Khaykovich}},\ }%
  \bibfield{journal}{%
  \bibinfo {journal} {Physical review letters}\ }%
  \textbf{\bibinfo {volume} {103}},\ \bibinfo {pages} {163202} (\bibinfo {year}
  {2009})%
  \bibAnnoteFile{NoStop}{loss_lithium_gross}%
\bibitem{loss_sodium_stenger}%
  \BibitemOpen
  \bibfield{author}{%
  \bibinfo {author} {\bibfnamefont{J.}~\bibnamefont{Stenger}}, \bibinfo
  {author} {\bibfnamefont{S.}~\bibnamefont{Inouye}}, \bibinfo {author}
  {\bibfnamefont{M.}~\bibnamefont{Andrews}}, \bibinfo {author}
  {\bibfnamefont{H.-J.}\ \bibnamefont{Miesner}}, \bibinfo {author}
  {\bibfnamefont{D.}~\bibnamefont{Stamper-Kurn}},\ and\ \bibinfo {author}
  {\bibfnamefont{W.}~\bibnamefont{Ketterle}},\ }%
  \bibfield{journal}{%
  \bibinfo {journal} {Physical review letters}\ }%
  \textbf{\bibinfo {volume} {82}},\ \bibinfo {pages} {2422} (\bibinfo {year}
  {1999})%
  \bibAnnoteFile{NoStop}{loss_sodium_stenger}%
\end{thebibliography}%
  
\end{document}